\documentclass[useAMS,usenatbib]{mn2e}
\usepackage{graphicx,amssymb,times}

\title[7-mm spectral imaging of the CMZ]{Spectral imaging of the Central 
Molecular Zone in multiple 7-mm molecular lines}
\author[P. A. Jones et al.]{P. A. Jones$^{1}$
\thanks{E-mail:pjones@phys.unsw.edu.au},
M. G. Burton$^{1}$, M. R. Cunningham$^{1}$, N. F. H. Tothill$^{1,2}$, A. J. 
Walsh$^{3}$
\\
$^{1}$ School of Physics, University of New South Wales, NSW 2052, Australia \\ 
$^{2}$ University of Western Sydney, Locked Bag 1797, Penrith South DC, NSW 
1797, Australia \\
$^{3}$ International Centre for Radio Astronomy Research, Curtin University, 
Perth, WA, 6845, Australia \\
}

\begin{document}

\date{Accepted . Received ; in original form 2012 April 1}

\pagerange{\pageref{firstpage}--\pageref{lastpage}} \pubyear{2013}

\maketitle

\label{firstpage}

\begin{abstract}

We have imaged 24 spectral lines in the Central Molecular Zone
(CMZ) around the Galactic Centre, in the range 42 to 50~GHz.  The
lines include emission from the CS, CH$_{3}$OH, HC$_{3}$N, SiO, HNCO,
HOCO$^{+}$, NH$_{2}$CHO, OCS$^{+}$, CCS, C$^{34}$S, $^{13}$CS, $^{29}$SiO,
H$^{13}$CCCN, HCC$^{13}$CN and HC$_{5}$N
molecules, and three hydrogen recombination lines.
The area covered is Galactic longitude $-0.7$ to $1.8$~deg. and latitude 
$-0.3$ to $0.2$~deg., including the bright cores around Sgr~A, Sgr~B2, 
Sgr~C and G1.6-0.025.   This
work used the 22-m Mopra radio telescope in Australia, 
obtaining $\sim 1.8$~km~s$^{-1}$ spectral and $\sim 65$~arcsec spatial resolution.
We present peak images from this study and conduct a
  principal component analysis on the integrated emission from the
  brightest 10 lines, to study similarities and differences in the line distribution.
We examine the integrated line intensities and line ratios in selected apertures
  around the bright cores, as well as for the complete mapped
  region of the CMZ. 
We compare these 7-mm lines to the corresponding lines in the 3-mm band,
for five molecules, to study the excitation. There is a variation
in 3-mm to 7-mm line ratio across the CMZ, with relatively higher ratio
in the centre around Sgr~B2 and Sgr~A. We find that the lines are
sub-thermally excited, and from modelling with {\sc radex} find that
non-LTE conditions apply, with densities of order 10$^{4}$ cm$^{-3}$.
\end{abstract}


\section{Introduction}
\label{sec:intro}

The Central Molecular Zone (CMZ) is the region within a few hundred parsecs
of the Galactic Centre \citep{mose96} with dense molecular gas, of total 
molecular mass a few times $10^{7}$~M$_{\sun}$ \citep*{fegije07}. The CMZ is an 
unusual environment, with high molecular gas densities, large velocity
dispersion and shears, high temperature, and rich chemistry: see
\citet{jo+12} and references therein.

Although there is on-going star formation in the CMZ, the 
star-formation rate (SFR) is an order of magnitude
less than would be expected from
such a large amount of molecular gas \citep{lo+13} using common SFR scaling 
relations. Hence, studying the properties of the molecular medium in the CMZ is 
important in understanding the enviromental dependence of star formation,
particularly the environments of the centres of galaxies.

We present here observations of the CMZ of 24 lines in the 7-mm band (including
three hydrogen recombination lines),
with resolution around 65 arcsec. 

This is part of a larger project in which have also imaged the CMZ area
in multiple lines in the 3-mm band, with resolution around 40 arcsec
\citep{jo+12}, and in the 12-mm band,
with resolution around 120 arcsec, as part of the wider H$_{2}$O southern 
Galactic Plane Survey \citep[HOPS,][]{wa+08,wa+11, pu+12}. 
A further Mopra survey to image CO lines over a larger area of the CMZ is
ongoing. 
We have also studied the smaller Sgr~B2 
sub-area of the CMZ with the molecular lines
over the full 3-mm band from 82 to 114~GHz,
and 7-mm band from 30 to 50~GHz, in \citet{jo+08} and \citet{jo+11}
respectively.

We use the distance to the Galactic Centre of 8.0~kpc \citep{re+09}, 
as per our previous papers, so 1~arcmin corresponds to 2.3~pc.

\section[]{Observations and Data Reduction}
\label{sec:obs}

The observations were made with the 22-m Mopra radio telescope, of the 
Australia Telescope National Facility (ATNF) using the UNSW 
MOPS digital filterbank.
The Mopra MMIC (Monolithic Microwave Integrated Circuit) receiver has a 
bandwidth of 8~GHz, and the MOPS backend can
cover the full 8-GHz range simultaneously in the broad band mode. This gives 
four 2.2-GHz sub-bands each with 8192 channels of 0.27~MHz. 
The lines in the CMZ are broad, so that 
these channels, corresponding to around 1.8~km~s$^{-1}$, are quite adequate. 

The Mopra receiver covers the range 30 to 50~GHz in the 7-mm band. We chose
the tuning centred at 45.8~GHz, to cover the range 41.8 to 49.8~GHz, giving the
spectral lines summarised in Table \ref{tab:lines_table}. This range was chosen
to include the strong lines of CS near 49.0 GHz and SiO near 43.4 GHz, 
and several other bright lines as listed in the Table. This is also the
densest part of the 30 to 50~GHz band for bright molecular lines, as seen
by \citet{jo+11} in Sgr~B2.

\begin{table}
\begin{center}
\caption{The lines imaged here in the 42 to 50~GHz range.  
The rest frequencies with a * indicate the frequency used for lines
corresponding to multiple transitions, some also 
indicated as group (gp) in the transition list. The multiple
components are, in general, blended due the large velocity
widths in the CMZ area. The Mopra observations covered the whole 42 to 50~GHz 
range, so include weaker lines than are listed here, particularly in 
Sgr~B2, see \citet{jo+11}.}
\label{tab:lines_table}
\begin{tabular}{cccc}
\hline
Rough     & Line ID          & Transition         & Exact      \\
frequency & molecule         &                    & rest frequency \\
GHz       & or atom          &                    & GHz        \\
\hline
42.39  &  NH$_{2}$CHO   &  2(0,2)~--~1(0,1) gp    &  42.386070*   \\  
42.60  &  HC$_{5}$N     &  16~--~15  	          &  42.602153    \\
42.67  &  HCS$^+$       &  1~--~0                 &  42.674197    \\
42.77  &  HOCO$^+$      &  2(0,2)~--~1(0,1)       &  42.7661975   \\ 
42.82  &  SiO           &  1~--~0 v=2             &  42.820582    \\
42.88  &  $^{29}$SiO    &  1~--~0 v=0             &  42.879922    \\
42.95  &  H             &  53 $\alpha$            &  42.951968    \\ 
43.12  &  SiO           &  1~--~0 v=1             &  43.122079    \\
43.42  &  SiO           &  1~--~0 v=0  	          &  43.423864    \\
43.96  &  HNCO          &  2(0,2)~--~1(0,1) gp    &  43.962998*   \\
44.07  &  CH$_{3}$OH    &  7(0,7)~--~6(1,6) A++   &  44.069476    \\       
44.08  &  H$^{13}$CCCN  &  5~--~4                 &  44.084172    \\    
45.26  &  HC$_{5}$N     &  17~--~16               &  45.264720    \\     
45.30  &  HC$^{13}$CCN  &  5~--~4                 &  45.297346    \\         
       &  HCC$^{13}$CN  &  5~--~4                 &  45.301707*   \\
45.38  &  CCS           &  4,3~--~3,2             &  45.379029    \\
45.45  &  H             &  52 $\alpha$            &  45.453719    \\
45.49  &  HC$_{3}$N     &  5~--~4 gp              &  45.490316*   \\
46.25  &  $^{13}$CS     &  1~--~0                 &  46.247580    \\
47.93  &  HC$_{5}$N     &  18~--~17               &  47.927275    \\      
48.15  &  H             &  51 $\alpha$            &  48.153597    \\
48.21  &  C$^{34}$S     &  1~--~0                 &  48.206946    \\
48.37  &  CH$_{3}$OH    &  1(0,1)~--~0(0,0) A++   &  48.372467*   \\      
       &  CH$_{3}$OH    &  1(0,1)~--~0(0,0) E     &  48.376889    \\
48.65  &  OCS           &  4~--~3                 &  48.6516043   \\         
48.99  &  CS            &  1~--~0                 &  48.990957    \\

\hline
\end{tabular}
\end{center}
\end{table}

The area was observed with on-the-fly (OTF) mapping, in a similar way to that 
described in \citet{jo+12}, for the 3-mm CMZ observations.

The $10 \times 10$ arcmin$^{2}$ OTF maps used a scan rate of 
8~arcsec/second, with 24~arcsec line spacing,
taking a little over an hour per block. We made pointing observations
of SiO maser positions (AH Sco or VX Sgr), after every second map, to correct 
the pointing to within 
about 10 arcsec accuracy.  The system temperature was calibrated 
with a continuous noise diode, but the 7-mm calibration system (unlike the 
3-mm system) does not include ambient-temperature paddle scans, so
some extra steps in processing were needed (see below) to correct for the 
atmospheric opacity. We included observations of line sources (Sgr~B2, M17SW
and G345.5+1.5) a few times each day, under different weather conditions
and at different elevations, to check the calibration.

We used position switching for bandpass calibration with
the off-source reference position observed before each 10 arcmin long source
scan. The reference position
(($\alpha, \delta)_{\rm J2000} = 17^{\rm h}51^{\rm m}03\rlap.$\,$^{s}$\,$6, 
-28^{\circ}22'47''$, or
$l = 1.093$ deg., $b = -0.735$ deg.) was carefully selected to be relatively 
free of emission.

Each $10 \times 10$ arcmin$^{2}$ block was observed twice, with Galactic
latitude and longitude scan directions, 
to reduce scanning direction stripes and to 
improve the signal to noise. The overall $2.5 \times 0.5$ deg$^{2}$ area was 
covered with a $15 \times 3$ grid of blocks, separated by 10~arcmin steps, 
making 90 OTF observations
required. In practice a few more OTF observations were used, including some
areas with OTF maps stopped by weather or other problems and re-observed,
or just areas re-observed in better weather.
The area covered is between $-0.72$ to $1.80$ degrees Galactic longitude, and 
$-0.30$ to $0.22$ degrees Galactic latitude.

The observations were spread over two periods, in March 2010
and 2011. The conditions,
in the southern hemisphere autumn, are warmer and
wetter than the best winter observing periods.
Although the blocks worst affected by poor weather were re-observed,
the range of conditions during the different observations means that there
are inevitably some blocks with greater noise level than others. In practice,
we observed the central area including Sgr~B2 and Sgr~A in 2010, and the outer
areas in 2011. The weather in 2011 Mar was not as good as in 2010, so that the
system temperature was higher, and the noise level greater in the outer areas. 

The OTF data were turned into FITS data cubes with the {\sc livedata} and
{\sc gridzilla} 
packages\footnote{http://www.atnf.csiro.au/people/mcalabre/livedata.html},
in a similar way to \citet{jo+12}. The bandpass correction used the off-source
spectra, and a second order polynomial, and the gridding used a median filter.

The FITS cubes were then read into the {\sc miriad} package for further
processing and analysis. The velocity pixels were  0.27~MHz or 
$\sim$1.8~km~s$^{-1}$, but to improve the signal-to-noise, and ensure
the lines were Nyquist sampled, we applied a 3-point Hanning smoothing,
leading to an effective resolution 0.54~MHz or $\sim$3.6~km~s$^{-1}$.
This is quite adequate spectral resoution for the
broad lines in the CMZ, which are $> 10$~km~s$^{-1}$ wide, 
but for narrow spectral features, such as the maser lines, we use the
data cubes in the original unsmoothed versions.

Small residual baseline offsets in the spectra were fitted and removed with a 
first order polynomial using the {\sc miriad} task contsub.

The resolution of the Mopra beam varies between $1.14 \pm 0.05$~arcmin at
43~GHz to $0.99 \pm 0.04$~arcmin at 49~GHz \citep{ur+10}. We take this
as the beamsize inversely proportional to frequency, with $1.06$~arcmin or 
64~arcsec at 46~GHz in the centre of the frequency range observed.

The main beam efficiency of Mopra 
varies between $0.53 \pm 0.01$ at 43~GHz and $0.43 \pm 0.01$ at 49~GHz, 
and the extended beam efficiency between $0.69 \pm 0.01$ at 43~GHz and 
$0.56 \pm 0.01$ at 49~GHz \citep{ur+10}.
Since we are largely concerned in this paper with the spatial
and velocity structure, we have mostly left the intensities 
in the $T_{A}^*$ scale,
without correction for the beam efficiency onto the $T_{MB}$ or $T_{XB}$ scales,
except for some of the quantitative analysis sections.

A flux correction has been applied to the data cubes to take into account
the atmospheric attenuation. The test spectra (of Sgr~B2, M17SW
and G345.5+1.5) showed a significant (a few times 10~\%) decrease in flux
associated with higher system temperatures. This was consistent with the
effect expected with increased optical depth $\tau$ in the atmosphere, at lower
elevation or higher humidity, increasing $T_{sys}$ with the addition of term 
$T_{amb} (1 - \exp(-\tau)) \sim \tau T_{amb} $ 
(where $T_{amb} \sim$~290~K is the atmospheric 
temperature) and decreasing the flux by factor $\exp(-\tau)$.
The system temperatures were measured in each of the 2~GHz sub-bands, across
the overall 8~GHz range observed. When combined in the data cubes, averaged
over the two passes in latitude and longitude scanning directions,
the means were 84, 93, 106 and 126 K at around 43, 45, 47 and 49 GHz 
respectively, with RMS scatter 10, 11, 13 and 15 K respectively. However, due
to outliers of data taken in poor weather, the full ranges were 68 -- 190,
75 -- 201, 85 -- 219 and 101 -- 231 K respectively.  
This provides a measure of the average optical depth $\tau$ as a function of 
position
in the images, 
so allowing a correction for it to be applied. 
The flux correction factors
applied to the cubes, as a function of position derived from the $T_{sys}$ 
images were $1.150 \pm 0.044$, $1.180 \pm 0.054$, $1.233 \pm 0.068$ and 
$1.276 \pm 0.087$ in the four sub-bands. 

\begin{table*}
\begin{center}
\caption{Statistics of the data cubes.  Both the RMS noise and 
peak brightness temperature are in $T_{A}^{*}$, after the calibration correction
for atmospheric attenuation. The velocity range
is that with line emission significantly above the noise level. 
We also quote the position and velocity of the peak pixel.}
\label{tab:summary_table}
\begin{tabular}{cccccccc}
\hline
Line      & Molecule         & RMS       & Velocity & Peak & \multicolumn{2}{c}{Peak} 
& Peak \\
Freq.     & or atom ID       & noise     & range  &        & lat. & long. & vel. \\
GHz       &                  & mK        & km~s$^{-1}$ & K & deg. & deg.  & 
km s$^{-1}$ \\
\hline
42.39  &  NH$_{2}$CHO   &   24 &    -5, 110  &  0.30  &  359.861 &  -0.082 &   8  \\  
42.60  &  HC$_{5}$N     &   25 &     0, 100  &  0.33  &    0.687 &  -0.020 &  65  \\
42.67  &  HCS$^+$       &   26 &    -5,  85  &  0.24  &    0.675 &  -0.020 &  71  \\
42.77  &  HOCO$^+$      &   27 &    -5, 100  &  0.39  &    0.694 &  -0.020 &  62  \\ 
42.82  &  SiO           &   26 &  -150, 200  &  3.22  &    0.550 &  -0.057 & -25  \\
42.88  &  $^{29}$SiO    &   27 &     0,  85  &  0.17  &    0.836 &  -0.248 &  49  \\
42.95  &  H 53 $\alpha$ &   29 &    35,  95  &  0.39  &    0.669 &  -0.037 &  64  \\ 
43.12  &  SiO           &   28 &  -150, 200  &  2.74  &    0.550 &  -0.057 & -26  \\
43.42  &  SiO           &   40 &  -140, 205  &  1.48  &  359.893 &  -0.076 &  15  \\
43.96  &  HNCO          &   38 &  -160, 185  &  2.68  &    1.649 &  -0.053 &  51  \\
44.07  &  CH$_{3}$OH    &   30 &   -60, 195  &  3.37  &  359.986 &  -0.069 &  47  \\  
44.08  &  H$^{13}$CCCN  &   30 &     0,  90  &  0.22  &    0.698 &  -0.028 &  61  \\    
45.26  &  HC$_{5}$N     &   30 &     0,  80  &  0.32  &    0.690 &  -0.026 &  63  \\     
45.30  &  HCC$^{13}$CN  &   29 &    30, 105  &  0.25  &    0.675 &  -0.020 &  68  \\ 
45.38  &  CCS           &   31 &    -5, 100  &  0.35  &    0.687 &  -0.020 &  70  \\
45.45  &  H 52 $\alpha$ &   38 &    30,  90  &  0.42  &    0.669 &  -0.037 &  63  \\
45.49  &  HC$_{3}$N     &   37 &  -150, 200  &  2.67  &  359.981 &  -0.078 &  16  \\
46.25  &  $^{13}$CS     &   34 &   -15, 110  &  0.49  &  359.983 &  -0.071 &  52  \\
47.93  &  HC$_{5}$N     &   73 &    20,  80  &  0.37  &    0.675 &  -0.022 &  67  \\      
48.15  &  H 51 $\alpha$ &   67 &    40,  85  &  0.45  &    0.672 &  -0.035 &  58  \\
48.21  &  C$^{34}$S     &   59 &   -25, 140  &  0.81  &  359.982 &  -0.073 &  51  \\
48.37  &  CH$_{3}$OH    &   76 &  -160, 205  &  3.76  &    0.689 &  -0.024 &  71  \\
48.65  &  OCS           &   55 &   -10, 105  &  0.72  &    0.691 &  -0.029 &  64  \\  
48.99  &  CS            &   75 &  -185, 210  &  3.51  &  359.984 &  -0.069 &  54  \\
\hline
\end{tabular}
\end{center}
\end{table*}

\section[]{Results}
\label{sec:results}

\subsection{Line data cubes}
\label{subsec:cubes}

The frequencies and line identifications for the 24 strongest lines in the 
41.8 to 49.8~GHz range, are listed in Table \ref{tab:lines_table}.
These data are taken from the  NIST online
catalogue\footnote{http://physics.nist.gov/PhysRefData/Micro/Html/contents.html}
of lines known in the interstellar medium \citep{lodr04} and the splatalogue 
compilation\footnote{http://www.splatalogue.net/}.

Some statistics of the data cubes are listed in Table \ref{tab:summary_table}. 
The RMS noise
ranges from 24 to 76 mK, increasing with frequency across the 8~GHz range,
with the highest noise for the 47.93~GHz
HC$_{5}$N line, near the 47.8~GHz edge of the sub-band.

The velocity range listed in Table \ref{tab:summary_table} indicates the range
of significant emission (roughly greater than 3~$\sigma$) detected in the 
data cubes. The line widths in
the CMZ are large, and there is a large velocity gradient across the CMZ
(in the deep potential
well of the Galactic Centre). Hence the integrated emission needs to summed
over a large velocity range. This makes the integrated emission quite
sensitive to low level baselevel offsets which are small in terms of 
brightness (e.g. of order 10 mK) but become significant (e.g. of order 
K~km~s$^{-1}$) when integrated over the velocity range (order 100~km~s$^{-1}$).
The 7-mm Mopra data here have better baselevel stability than the 3-mm data of
\citet{jo+12}, due to the lesser sensitivity to the weather at lower frequencies.
However, the baseline problems do provide a limit, along with the thermal noise,
to the accuracy of the data. 
 
These data cubes are publicly available in the CSIRO-ATNF archive,
accessible from  http://atoa.atnf.csiro.au/CMZ, together with the 3-mm
and 12-mm data from the CMZ and Sgr~B2.\footnote{Further details may be also be 
obtained from the UNSW web site at 
www.phys.unsw.edu.au/mopracmz.}

\subsection{Peak emission images}
\label{subsec:peak_em}

\begin{figure*}
\includegraphics[angle=-90,width=16.8cm]{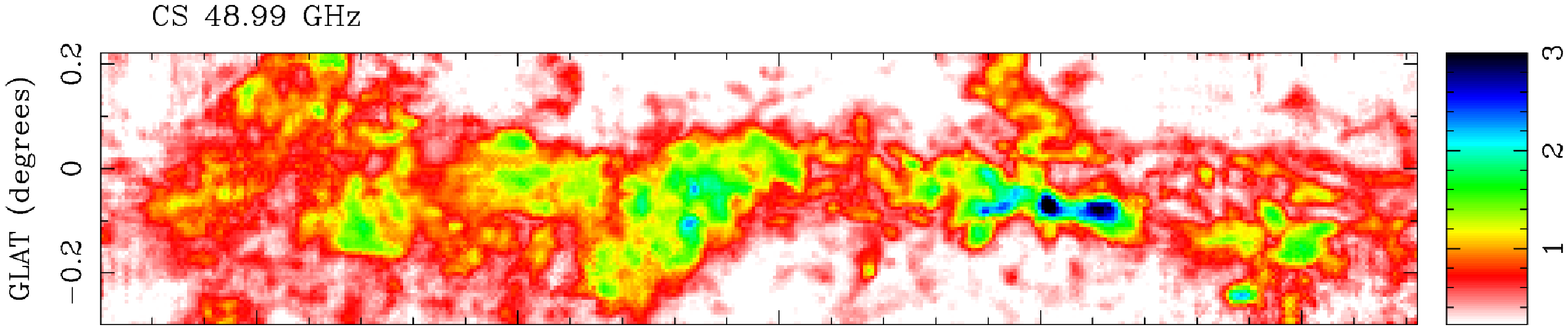}
\includegraphics[angle=-90,width=16.8cm]{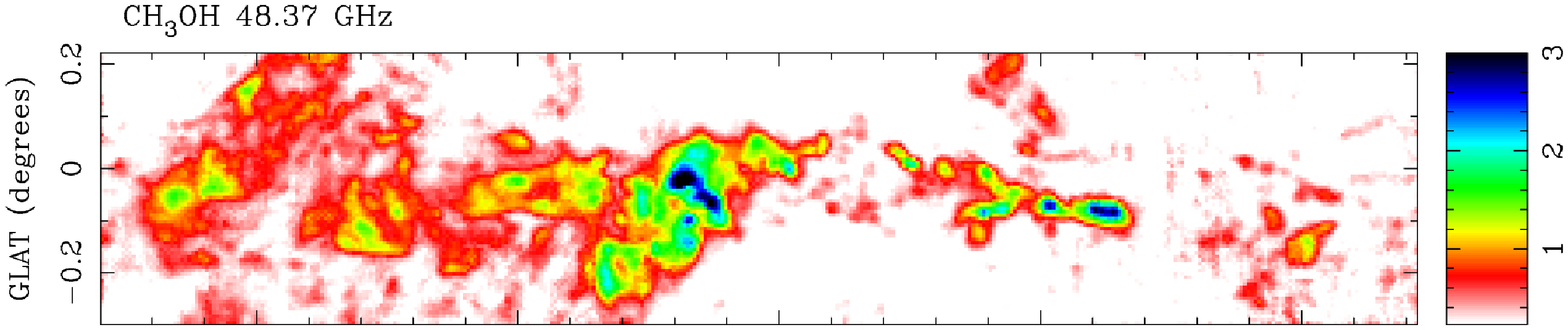}
\includegraphics[angle=-90,width=16.8cm]{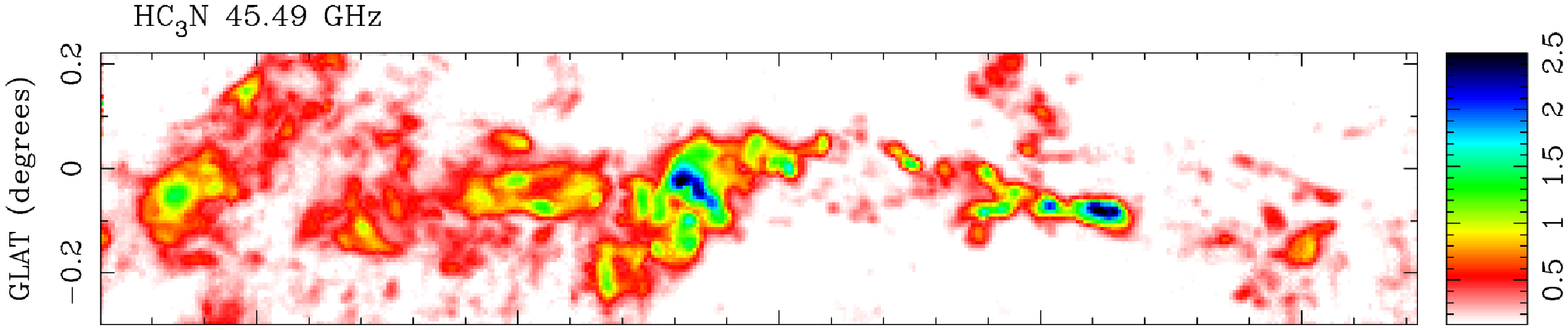}
\includegraphics[angle=-90,width=16.8cm]{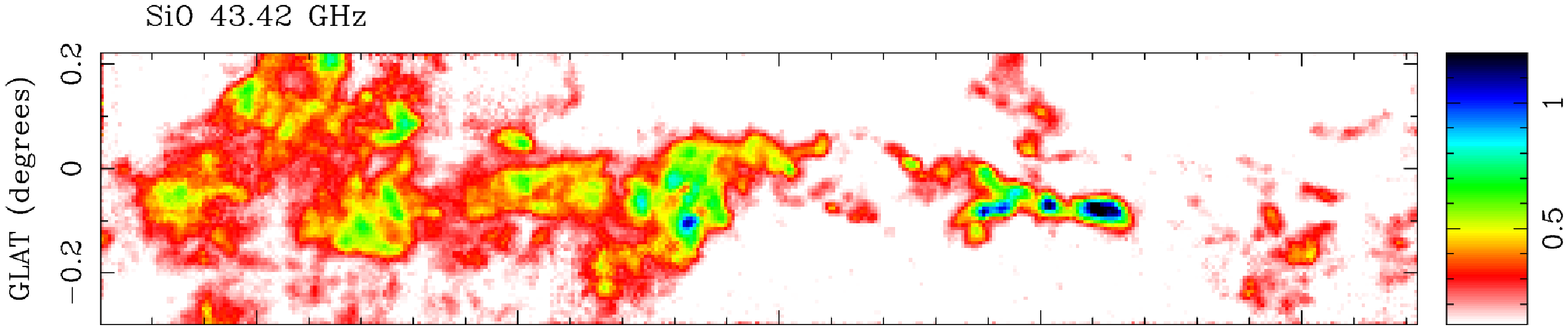}
\includegraphics[angle=-90,width=16.8cm]{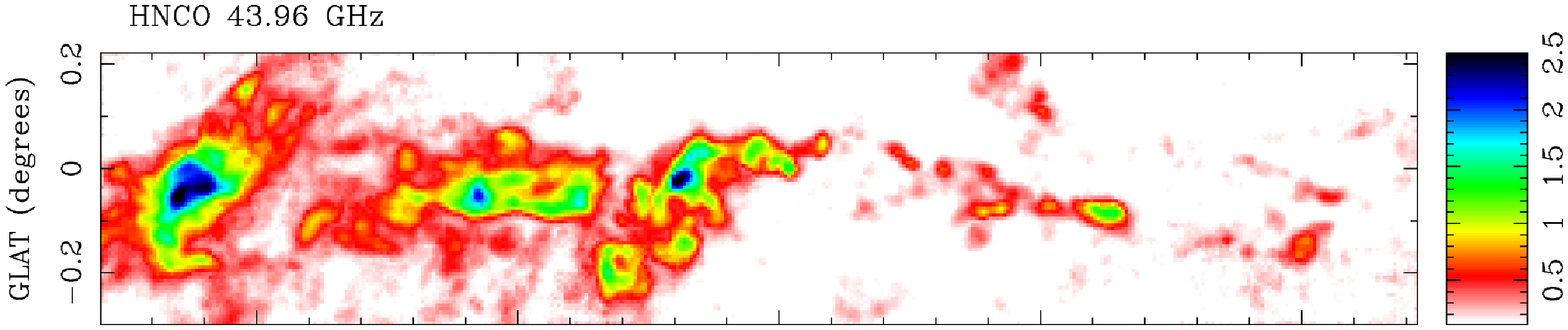}
\includegraphics[angle=-90,width=16.8cm]{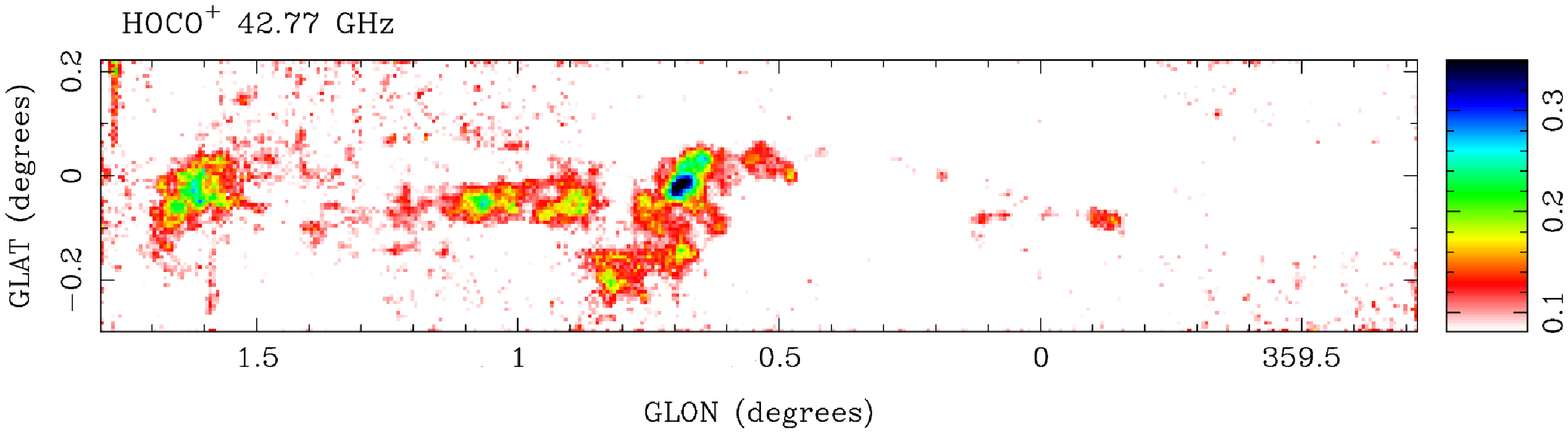}
\caption{Peak brightness images for the lines of CS, CH$_{3}$OH, HC$_{3}$N,
SiO, HNCO and HOCO$^{+}$.
These are the six lines we observed with the most widespread strong 
distribution.
The grey-scale is peak brightness as $T_{A}^{*}$ in K here and in Figs
\ref{fig:images_NH2CHO_etc} to \ref{fig:images_CH3OH_m_etc}.}
\label{fig:images_CS_etc}
\end{figure*}

\begin{figure*}
\includegraphics[angle=-90,width=16.8cm]{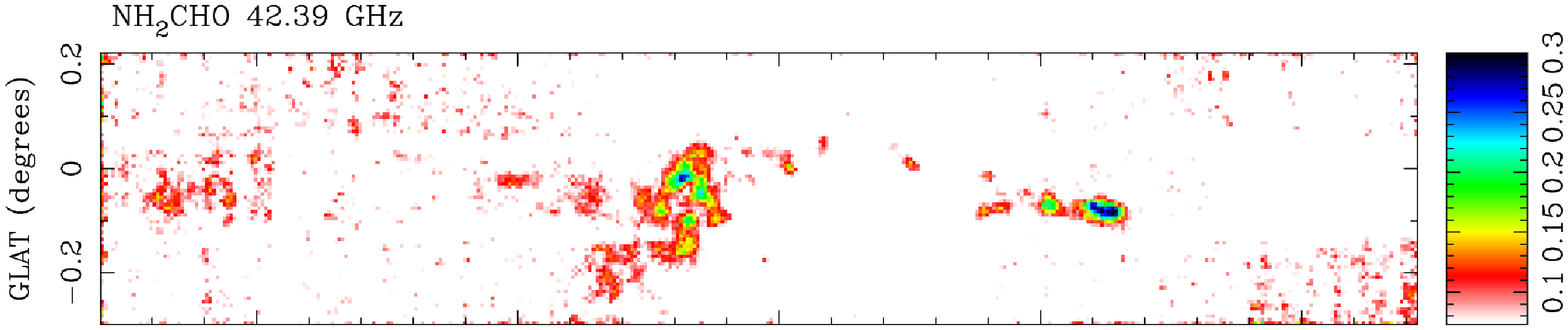}
\includegraphics[angle=-90,width=16.8cm]{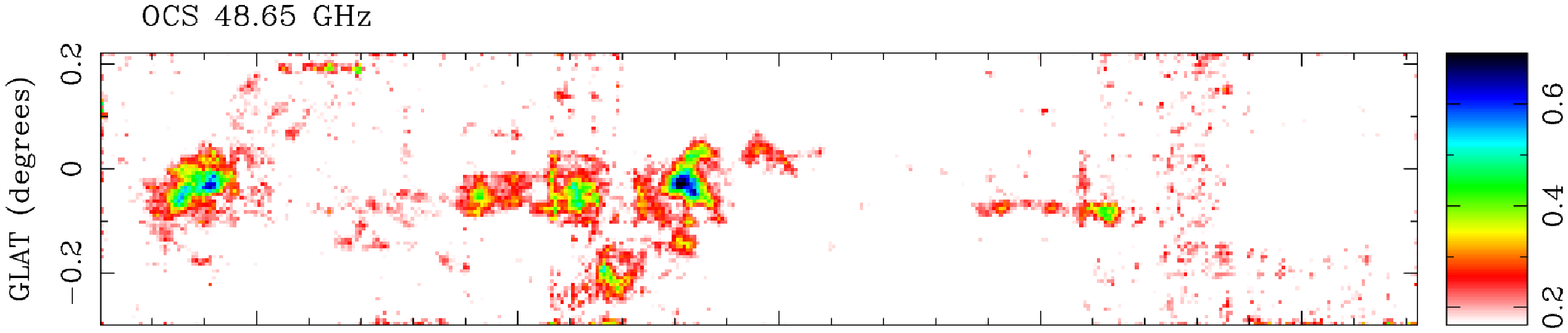}
\includegraphics[angle=-90,width=16.8cm]{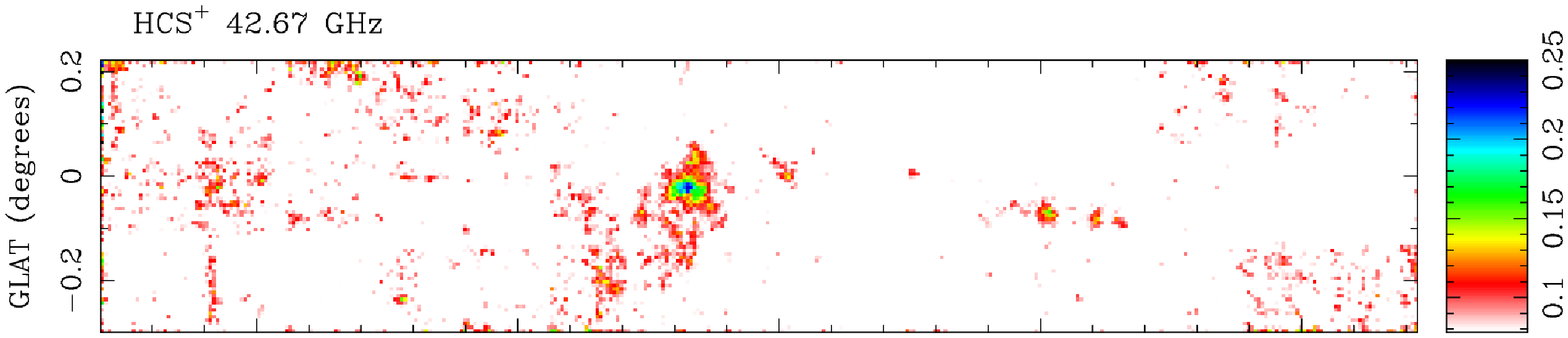}
\includegraphics[angle=-90,width=16.8cm]{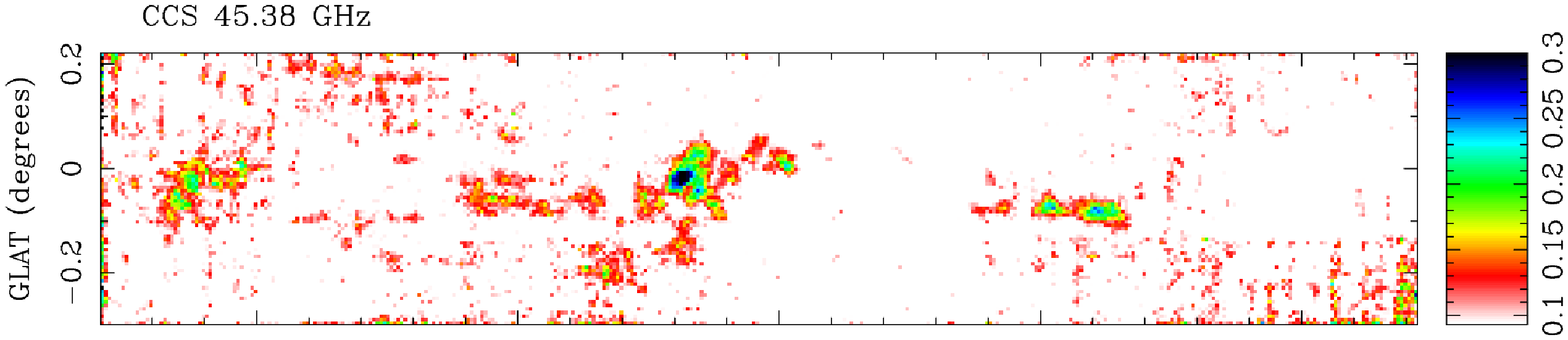}
\includegraphics[angle=-90,width=16.8cm]{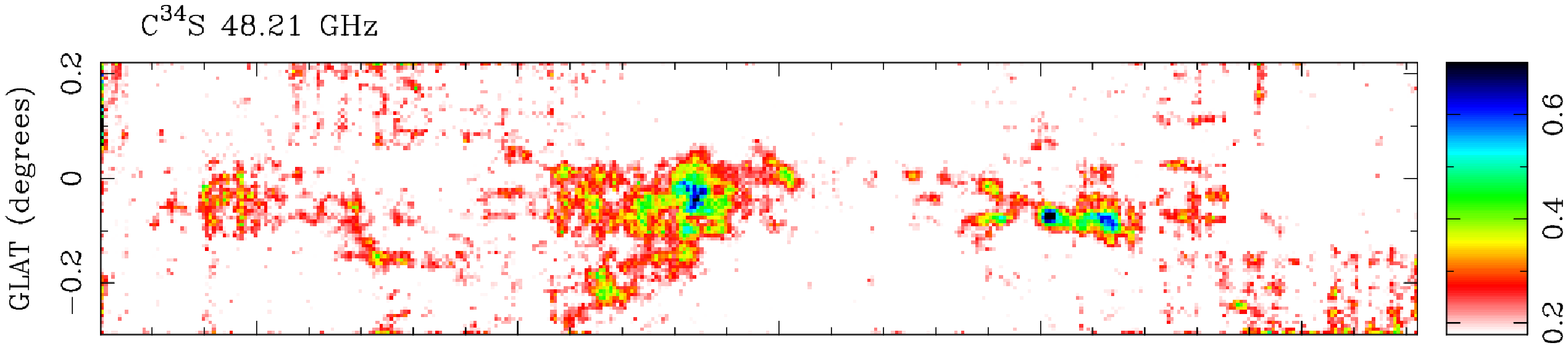}
\includegraphics[angle=-90,width=16.8cm]{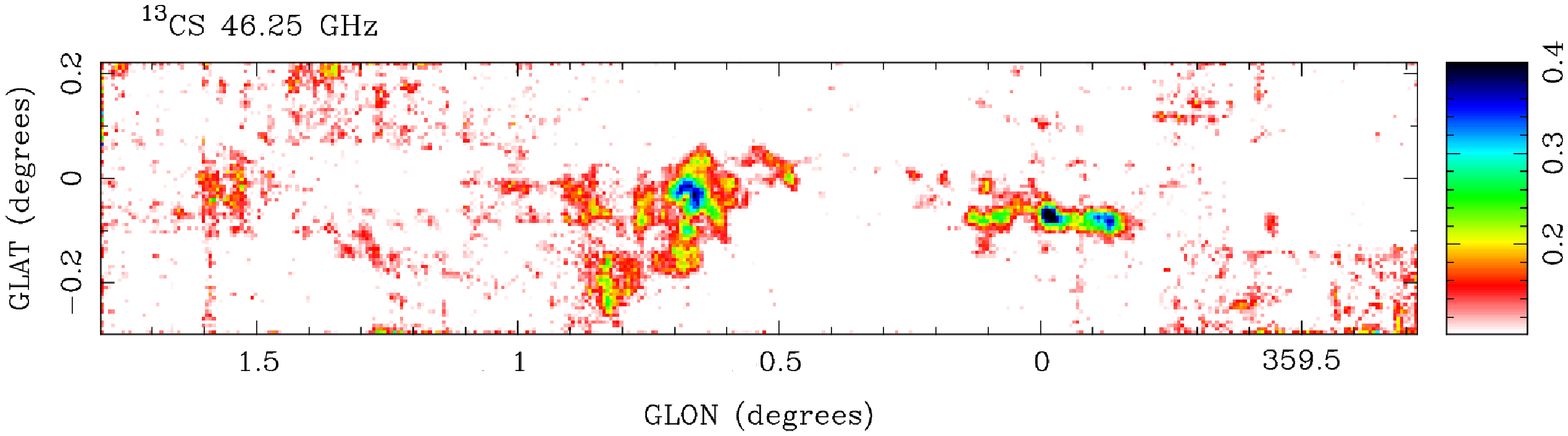}
\caption{Peak brightness images for the lines of NH$_{2}$CHO, OCS, HCS$^{+}$,
CCS and the CS isotopologues C$^{34}$S and $^{13}$CS.}
\label{fig:images_NH2CHO_etc}
\end{figure*}

\begin{figure*}
\includegraphics[angle=-90,width=16.8cm]{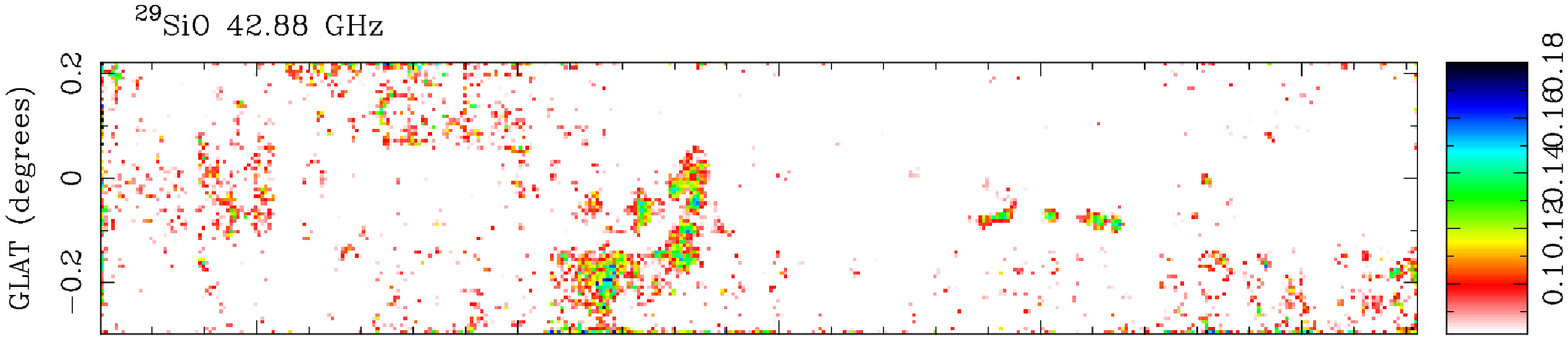}
\includegraphics[angle=-90,width=16.8cm]{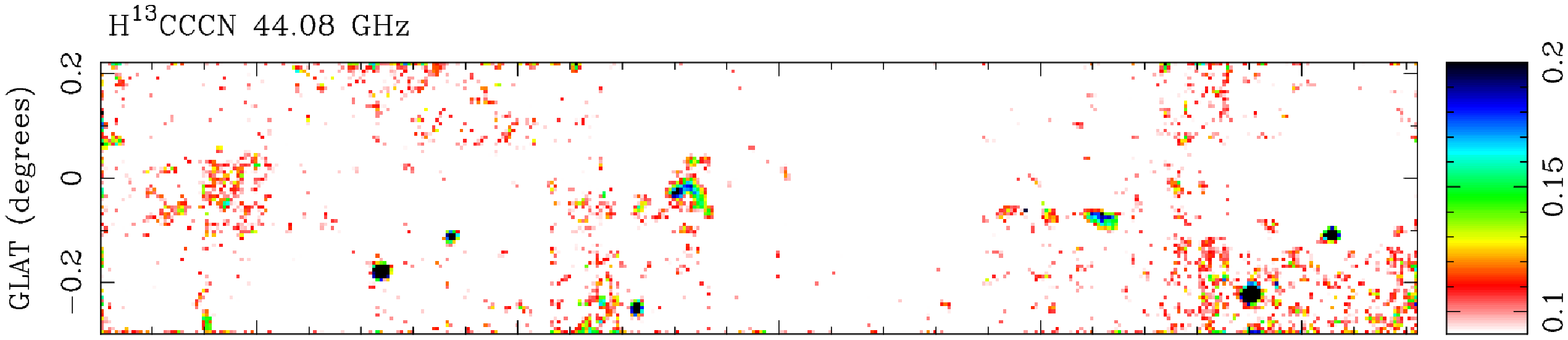}
\includegraphics[angle=-90,width=16.8cm]{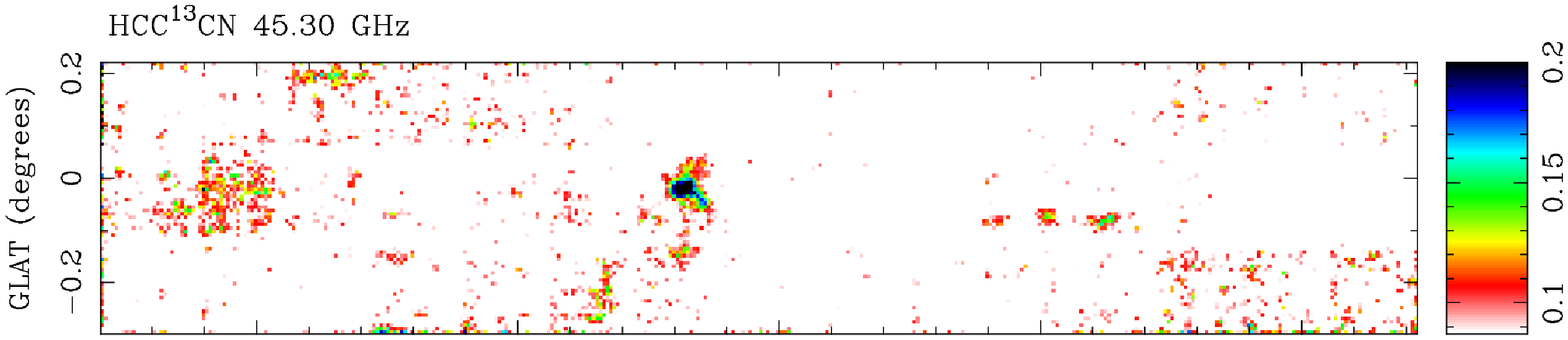}
\includegraphics[angle=-90,width=16.8cm]{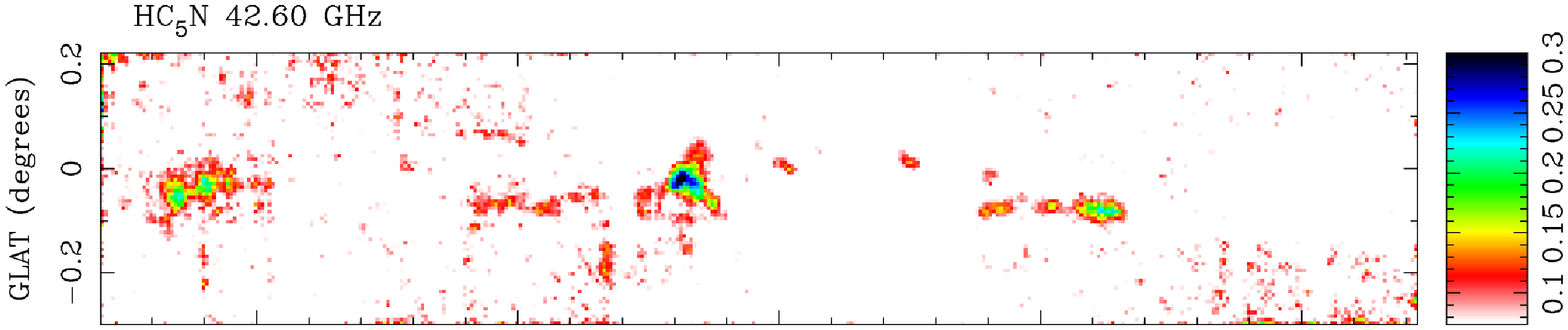}
\includegraphics[angle=-90,width=16.8cm]{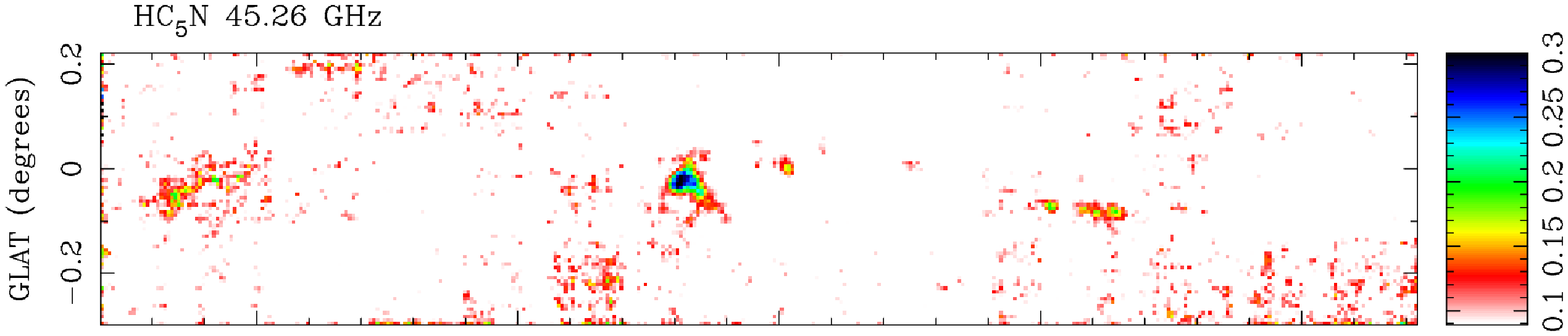}
\includegraphics[angle=-90,width=16.8cm]{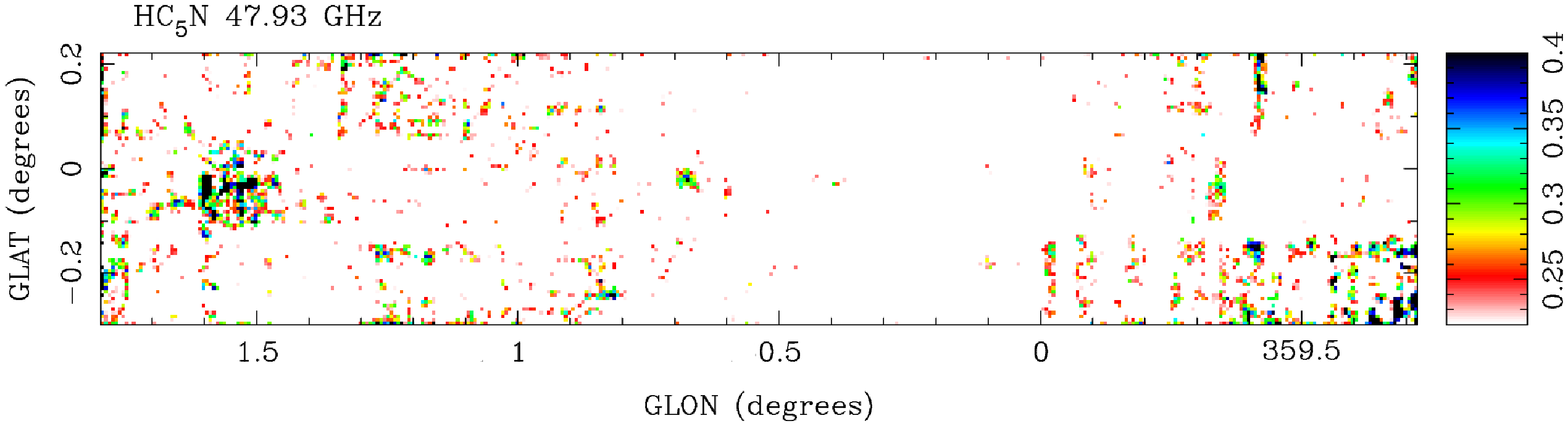}
\caption{Peak brightness images for the lines of SiO isotopologue $^{29}$SiO,
HC$_{3}$N isotopologues H$^{13}$CCCN and HCC$^{13}$CN blended with HC$^{13}$CCN,
and three transitions of HC$_{5}$N. Note that the 44.08~GHz H$^{13}$CCCN line is 
confused with the 44.07~GHz CH$_{3}$OH maser line (Fig. 
\ref{fig:images_CH3OH_m_etc}).}
\label{fig:images_29SiO_etc}
\end{figure*}

\begin{figure*}
\includegraphics[angle=-90,width=16.8cm]{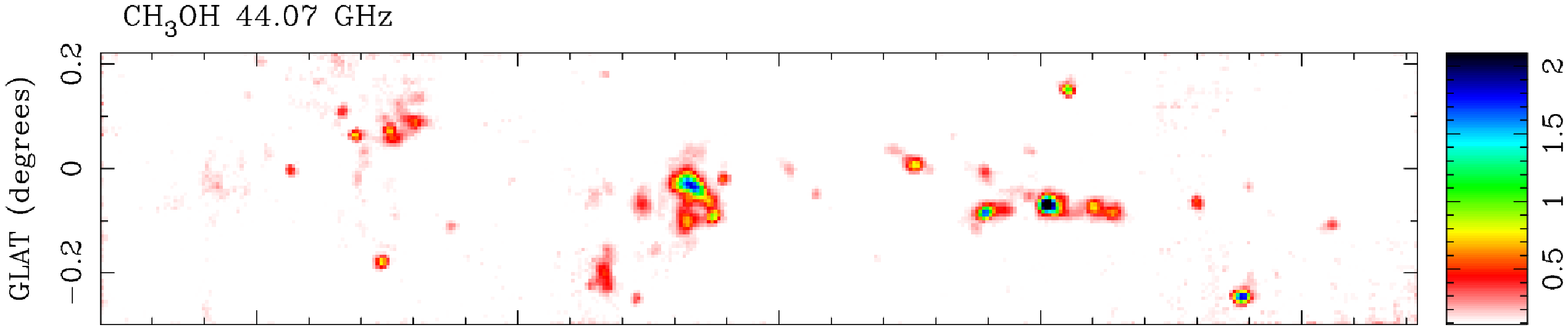}
\includegraphics[angle=-90,width=16.8cm]{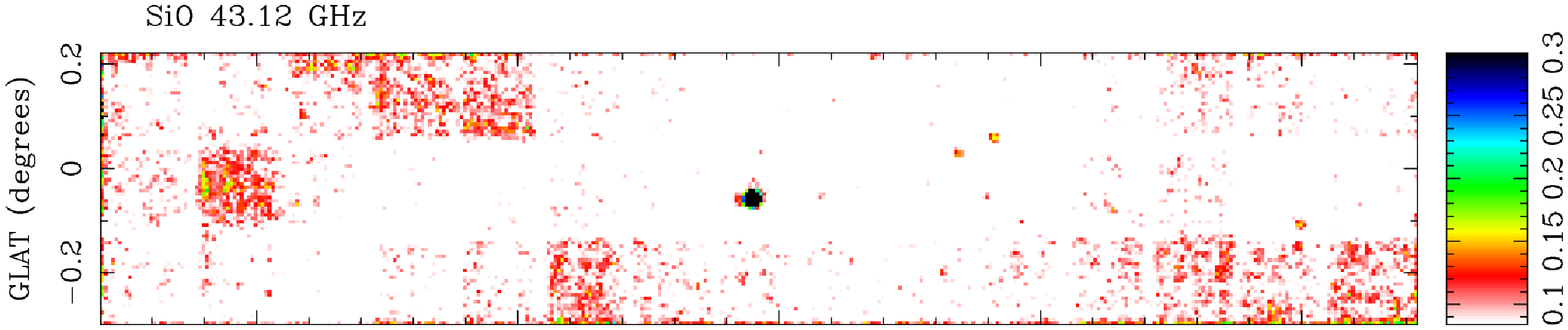}
\includegraphics[angle=-90,width=16.8cm]{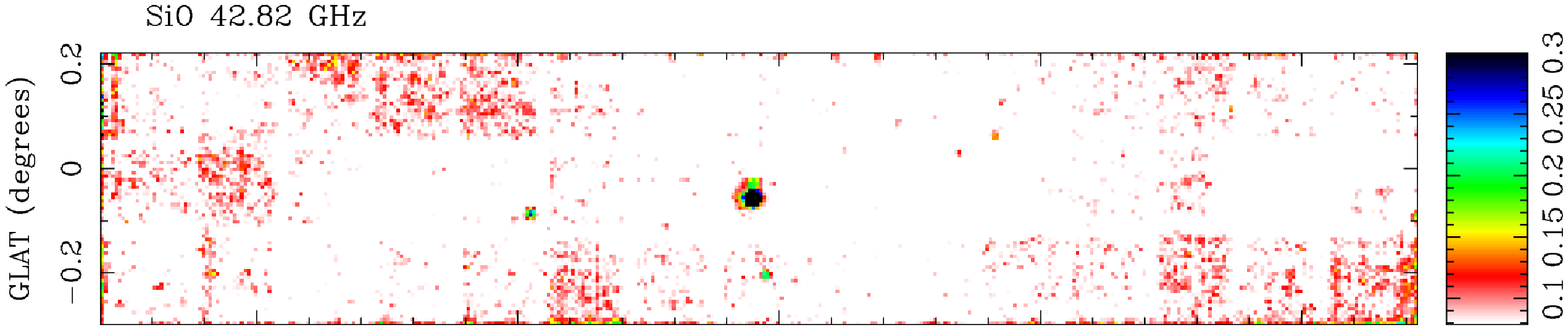}
\includegraphics[angle=-90,width=16.8cm]{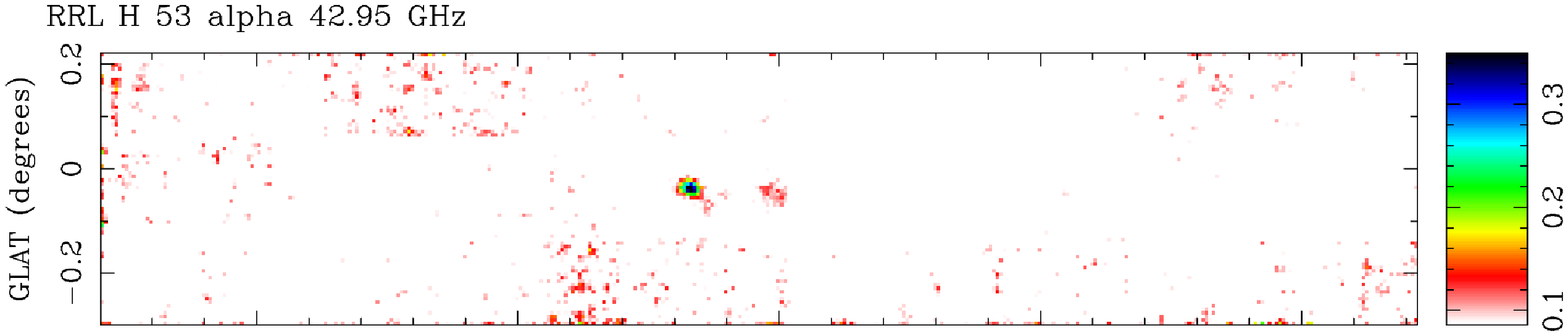}
\includegraphics[angle=-90,width=16.8cm]{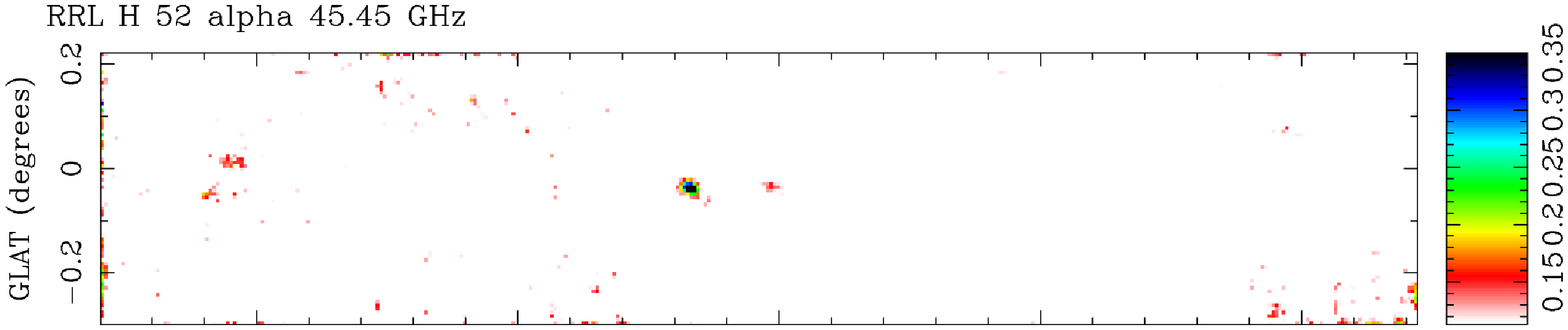}
\includegraphics[angle=-90,width=16.8cm]{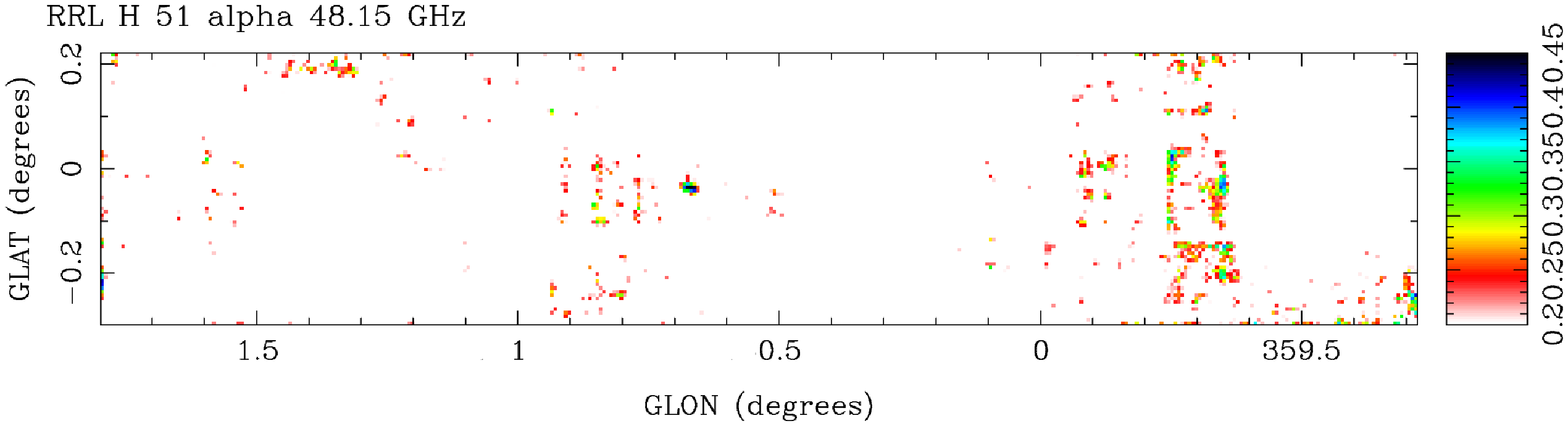}
\caption{Peak brightness images for lines of CH$_{3}$OH in a maser transition,
two maser transitions of SiO, and three hydrogen $\alpha$ recombination lines.}
\label{fig:images_CH3OH_m_etc}
\end{figure*}

We show the 
2-dimensional distribution of the line  
emission in Figs \ref{fig:images_CS_etc} to
\ref{fig:images_CH3OH_m_etc}. 
We use peak images as these are are much more robust to the effect of low 
level baseline problems than the integrated emission images. 
They also better show fine spatial structure as they are less affected
by multiple velocity components along the line of sight.
The colour-scale or grey-scale displays start at the 3~$\sigma$
brightness level, so do not show noisy features in areas without significant 
line emission. 
However, low-level emission outside the areas 
highlighted in these plots, may be found to be significant 
when the data are integrated over larger areas than the beam, or smoothed in 
velocity.

There are some artifacts in some of these peak brightness 
images, notably areas at longitude 359.6~deg. with higher noise, giving
maxima above the overall 3~$\sigma$ level.

The lines of CS, CH$_{3}$OH (at 48.37 GHz), HC$_{3}$N, SiO and HNCO
are relatively strong and widespread (Fig. \ref{fig:images_CS_etc}).
The lines of HOCO$^{+}$ (Fig. \ref{fig:images_CS_etc}), NH$_{2}$CHO, OCS, 
HCS$^{+}$, and CCS (Fig. \ref{fig:images_NH2CHO_etc}) are weaker,
and detected mostly in the Sgr~B2 and Sgr~A areas. The lines of HNCO, 
HOCO$^{+}$ and OCS highlight the core of Sgr~B2, and also the area of 
G1.6-0.025 (as discussed below in section \ref{subsec:PCA}).

The weaker isotopologues of CS, SiO and HC$_{3}$N, namely C$^{34}$S and 
$^{13}$CS in Fig. \ref{fig:images_NH2CHO_etc}, $^{29}$SiO, 
H$^{13}$CCCN and HCC$^{13}$CN (blended with HC$^{13}$CCN) in 
Fig. \ref{fig:images_29SiO_etc}, are also detected around the Sgr~B2 and 
Sgr~A areas. Note that the line of H$^{13}$CCCN at 44.08 GHz 
(Fig. \ref{fig:images_29SiO_etc}) is highly confused with the line of
CH$_{3}$OH at 44.07 GHz (Fig. \ref{fig:images_CH3OH_m_etc}), so the peak
image has additional strong CH$_{3}$OH maser peaks which overlap the frequency
range. In principle, these isotopologues could be used to measure isotope
ratios (C/$^{13}$C, S/$^{34}$ and Si/$^{29}$) near the Galactic Centre.
However, in practice, the low signal noise of the mapping data for these
weaker isotopolgues means that they are only detected at the strong cores,
where the strong isotopologues (CS, SiO and HC$_{3}$N) are
optically thick, so the line ratios do not give the intrinsic isotope ratios. 

The three lines of HC$_{5}$N (Fig. \ref{fig:images_29SiO_etc}) show emission 
from Sgr~B2, but the higher noise at the higher frequency means that the weaker
emission at Sgr~A and G1.6-0.025 detectable in the two lower frequency lines
is lost in the noise for the last line.

The line of CH$_{3}$OH at 44.07 GHz (Fig. \ref{fig:images_CH3OH_m_etc})
has maser as well as thermal emission, so shows diffuse thermal emission
around Sgr~B2 and Sgr~A, plus a number of maser spots distributed over a wider
area of the CMZ.

The two maser lines of SiO in Fig. \ref{fig:images_CH3OH_m_etc} are
dominated by a strong maser peak (at $l=$~0.550~deg., $b=$~-0.057~deg.)
but also show a number of weaker maser spots.
These are vibrationally excited lines, unlike the thermal v=0 line of SiO at 
43.42 GHz, and so have a very different distribution to the latter.

The three hydrogen $\alpha$ radio recombination lines are at Sgr~B2, which
is where the thermal bremsstrahlung emission in the CMZ is strongest.

More details on the 7-mm line distributions in the area around Sgr~B2
are given in \citet{jo+11}.

\subsection{Principal Component Analysis}
\label{subsec:PCA}

\begin{figure*}
\includegraphics[angle=-90,width=17.0cm]{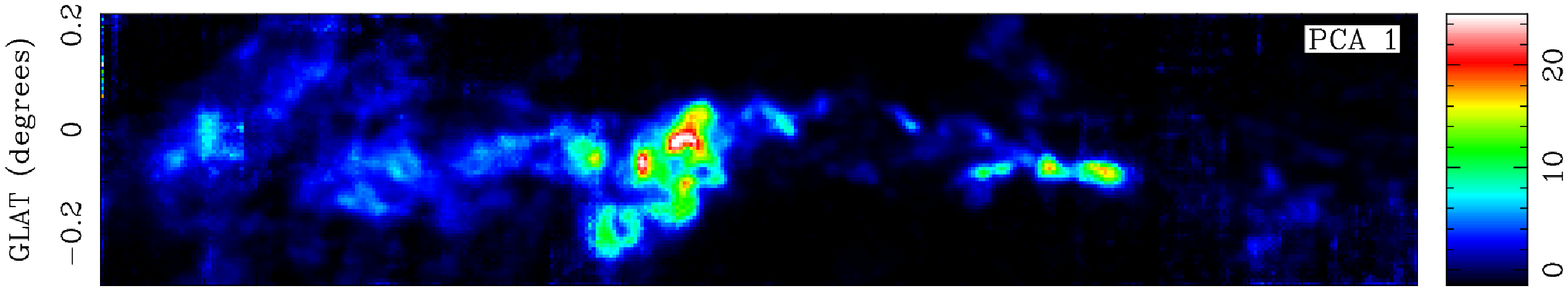}
\includegraphics[angle=-90,width=17.0cm]{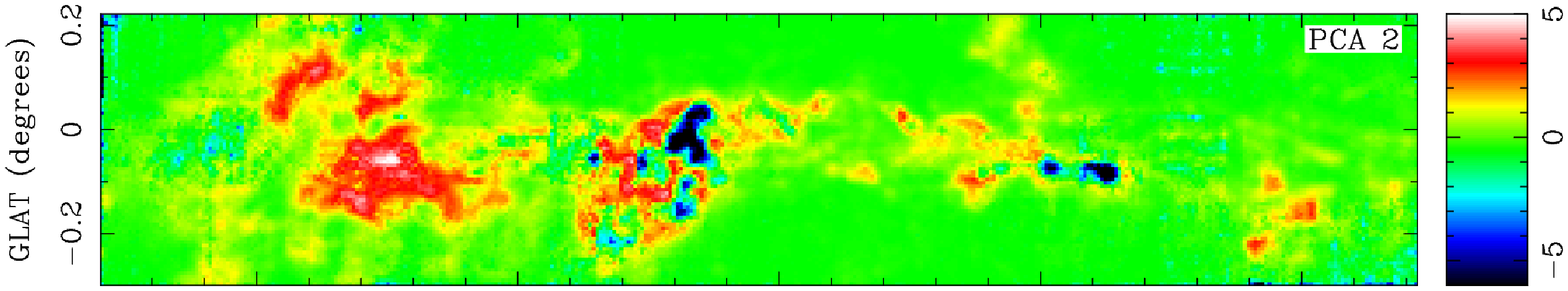}
\includegraphics[angle=-90,width=17.0cm]{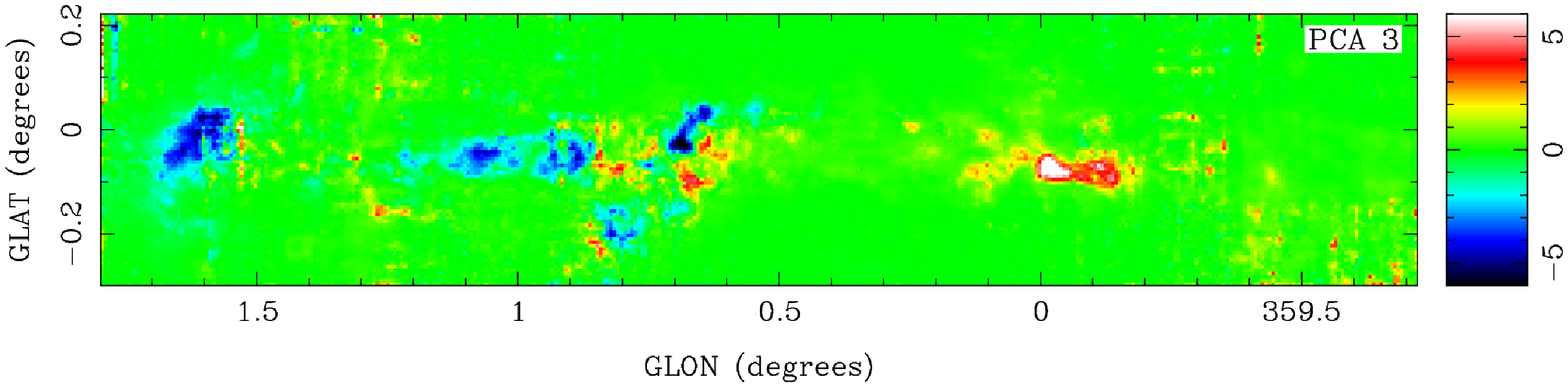}
\caption{The first three principal component images derived from the ten
strongest lines (CS, CH$_{3}$OH (at 48.37 GHz), HC$_{3}$N, SiO, HNCO, 
HOCO$^{+}$, NH$_{2}$CHO, OCS, C$^{34}$S and $^{13}$CS),
which describe 68.3, 11.5 and 7.3 percent of
the variance respectively.  The first
principal component describes the common features of the ten lines,
and the second principal component describes the most significant differences
between the lines. The further principal components describe successively
smaller differences. }
\label{fig:images_pca}
\end{figure*}

\begin{figure*}
\includegraphics[width=7cm]{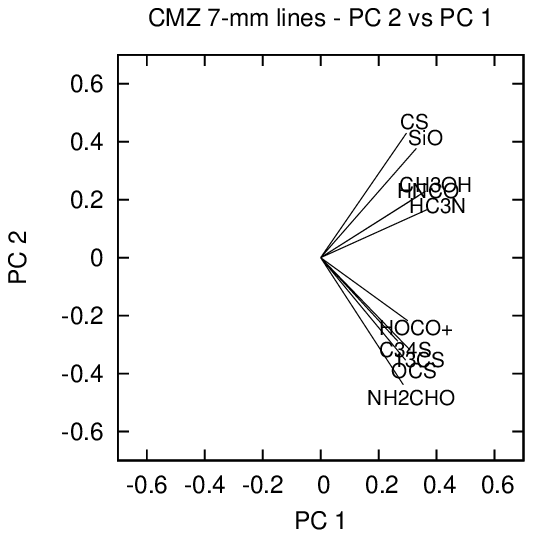}
\includegraphics[width=7cm]{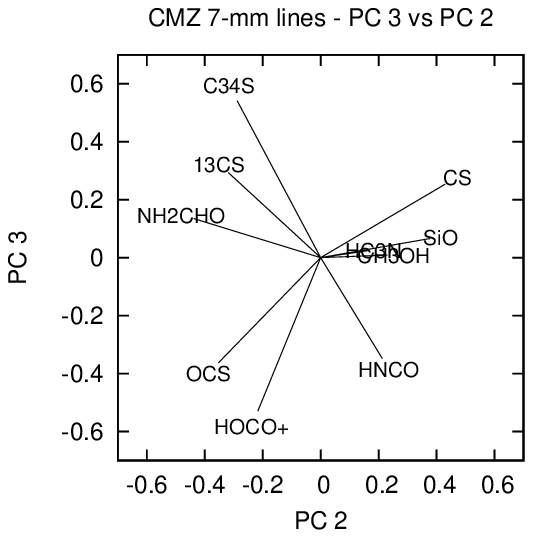}
\caption{The component vectors quantifying how the integrated line images are
composed of the sum of different scalar amounts of the of the principal component
images (Fig. \ref{fig:images_pca}).}
\label{fig:pca_vectors}
\end{figure*}

We can identify and quantify similarities and differences between the CMZ
line emission here with Principal Component Analysis (PCA). 
This technique describes
multi-dimensional data sets by linear combinations of the data that describe
the largest variance (the most significant common feature) and successively
smaller variances (the next most significant features). For 
integrated emission images here, PCA gives a series of images
which contain the most significant features (Fig. \ref{fig:images_pca})
and a set of coefficients (Fig. \ref{fig:pca_vectors}) which describe
how the individual line images are made up of linear combinations of
these PCA images.

We have used this PCA technique on the Mopra line surveys 
of the CMZ at 3-mm \citep{jo+12}, the Sgr~B2 area at 3-mm and 7-mm
\citep{jobulo08,jo+11}, and the G333 molecular cloud complex at 3-mm 
\citep{lo+09}. These papers, and the references therein, give more details on 
the technique.

Since the PCA uses normalised versions
of the input data sets, it does not work well with low signal to 
noise data, as the noisy data
is scaled up, causing the result to be dominated by spurious features.
Hence we use the ten strongest lines here (CS, CH$_{3}$OH (at 48.37 GHz), 
HC$_{3}$N, SiO, HNCO, HOCO$^{+}$, NH$_{2}$CHO, OCS, C$^{34}$S and $^{13}$CS)
out of the total of 24 lines. We consider the whole observed area, less
a small cropping at the low sensitivity edges of the area. This is different to
the PCA analysis of the CMZ 3-mm lines in \citet{jo+12}, where we applied an 
additional mask of low surface brightness areas, leaving only the stronger
areas around Sgr~B2 and Sgr~A, because the 3-mm data were more badly affected
by baseline ripples.

We use the versions of the integrated line
emission (over the velocity ranges given in Table \ref{tab:summary_table})
with clipping below the 3~$\sigma$ level, as these are less
affected by baseline ripple than the integrated emission without clipping.
However, we have checked the PCA results from integrated emission, with and 
without this clipping, and find qualitatively similar results. Hence the
PCA results are not too seriously affected by the small amount of missing
flux caused by the 3~$\sigma$-level clipping.

The images of the three most significant components of the integrated 
emission are shown in Fig. \ref{fig:images_pca}. These three components 
describe 68.3, 11.5 and 7.3 percent respectively 
of the variance in the data.
The normalised projection of the molecules onto the principal
components are shown in Fig. \ref{fig:pca_vectors} with successive pairs of
components.

The first principal component (Fig. \ref{fig:images_pca}) shows emission 
common to all ten lines. As all of these lines
are quite similar in overall morphology, it describes
a large fraction (68.3 percent) of the variance.
This behaviour is similar to that seen at 3-mm \citep{jo+12}, where the 
first component displays an averaged intensity image of all the molecular lines 
(albeit with a factor two better resolution).

The second principal component (Figs. \ref{fig:images_pca} and 
\ref{fig:pca_vectors})
shows the major difference (11.5  percent of the variance)
among the ten lines, which is the difference between the brightest regions
(blue in the colour version of Fig. \ref{fig:images_pca}, dark in the black and 
white version) and the fainter emission (red and yellow in the colour version,
light in the black and white version). We attribute this to the dense
cores having high optical depth in CS, SiO and (to a lesser extent) 
CH$_{3}$OH, HNCO and HC$_{3}$N, so that the peaks have less emission
than would be expected from optically thin lines (as the others are). This
is borne out by comparing specifically CS to the isotopologues 
C$^{34}$S and $^{13}$CS. This pattern of the second principal component
showing the optical depth effect is similar to that found in \citet{jo+12}
for the 3-mm CMZ lines. It is the case that this second principal component
is related to the strength of the line, so there could be some concern about
it being affected by the missing flux in using the integrated emission
with 3~$\sigma$ clipping, however (as noted above) the PCA using
the integrated emission without clipping gives similar results.

The third principal component shows the next most significant difference
among the ten lines (7.3 percent of the variance). This component shows
differences between the lines of CS, C$^{34}$S and $^{13}$CS on one side,
with a relative excess around Sgr~A (red and white in the colour version of 
Fig. \ref{fig:images_pca}, dark in the black and 
white version) and the lines of HOCO$^{+}$, HNCO and OCS on the other,
which show a relative excess around G1.6-0.025 and the peak of Sgr~B2
(blue in the colour version, light in the black and white version). 
The relative excess of HOCO$^{+}$, HNCO and OCS around G1.6-0.025 can
be seen by eye in Figs. \ref{fig:images_CS_etc} and \ref{fig:images_NH2CHO_etc}
(albeit plotted as peak emission, not integrated emission). 
HNCO has indeed been shown to vary significantly in relative intensity to dense 
gas tracers such as CS across the CMZ by \citet{ma+08} and \citet*{ammama11},
with the suggestion that this provides a diagnostic tool to distinguish 
between regions dominated by shock or by PDR emission from the molecules.
In the 3-mm PCA analysis \citep{jo+12} the shock-sensitive species HNCO and 
SiO were distinguished from the dense gas tracers.

The line differences in the Sgr~B2 area at 7-mm and 3-mm are shown in more
detail in  \cite{jo+11} and \cite{jobulo08}, and discussed there, but 
HOCO$^{+}$ and HNCO do highlight a chemically distinct region in the Sgr~B2 
area. 

We note that after the ``optical depth correction'' effect
of the second principal component that CS and weaker isotopologues 
C$^{34}$S and $^{13}$CS now show similar behaviour in the third principal 
component (same sign, $^{13}$CS and CS similar magnitude), giving more 
confidence that this third component is 
showing real differences in the molecular distributions.

We also note that the PCA images from the 7-mm lines here (Fig. 
\ref{fig:images_pca}) for the first three components agree very well with
the PCA images from the 3-mm lines in figure 9 of \citet{jo+12}, 
despite the fact that the latter are derived from different lines.
This gives confidence that the PCA technique is identifying physically
meaningful features. The sign of the principal components is arbitrary,
and the grey-scale display of figure 9 of \citet{jo+12} was chosen
differently to Fig. \ref{fig:images_pca} here, so the visual displays
of the two sets of PCA images are a bit different, but the data are quite 
similar.

\subsection{Line intensities and line ratios}
\label{subsec:areas_lines}

\begin{figure*}
\includegraphics[angle=-90,width=17cm]{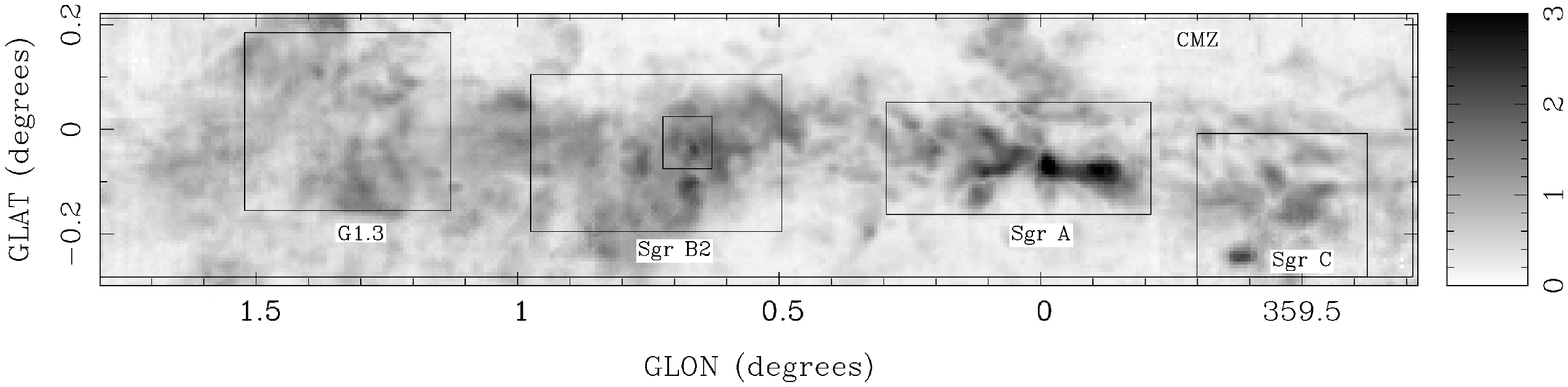}
\caption{The six areas selected for analysis: around Sgr~A, Sgr~B2, Sgr~C, G1.3,
a smaller area at the Sgr~B2 core, and the larger CMZ area, plotted on the peak
CS emission.}
\label{fig:areas}
\end{figure*}

The previous subsection (\ref{subsec:PCA}) indicates that there are differences
in the ratio of lines for different areas within the CMZ. We can quantify this
by integrating the line emission over different apertures.
The apertures chosen are listed in Table \ref{tab:apertures}
and plotted in Fig. \ref{fig:areas}. 
The six regions are the entire region of the CMZ that we have mapped, 
the four main emission features namely
Sgr~A, Sgr~B2, Sgr~C and G1.3, and a smaller region around the 
core of Sgr~B2.
They are effectively the same as those 
used in \citet{jo+12} for similar analysis of the CMZ in 3-mm lines, with very 
minor changes due to the data at 7-mm being on different (coarser) pixels
to the previous 3-mm data.

\begin{table}
\begin{center}
  \caption{Apertures selected for analysis. The areas are rectangles
    in Galactic coordinates, centred on $l$ and $b$. The 
    projected area on the sky assumes a distance of 8.0 kpc.}
\label{tab:apertures}
\begin{tabular}{ccccccccc}
\hline
Region     &   $l$   & $b$ & Width  & Height & Area         & Area   \\
           &   deg   & deg & arcmin & arcmin & arcmin$^{2}$ & pc$^2$ \\
\hline
CMZ         &  0.545 & -0.035  & 151.2 & 30.0 & 4\,536 & 24\,600 \\
Sgr~A       &  0.042 & -0.055  &  30.8 & 13.2 &    406 &  2\,200 \\
Sgr~B2      &  0.735 & -0.045  &  29.2 & 18.4 &    537 &  2\,910 \\
Sgr~C       & -0.462 & -0.145  &  20.0 & 16.8 &    336 &  1\,820 \\
G1.3        &  1.325 &  0.015  &  24.0 & 20.8 &    499 &  2\,700 \\
Sgr~B2      &  0.675 & -0.025  &   6.0 &  6.4 &   38.4 &    208  \\
Core        &        &         &       &      &        &         \\
\hline
\end{tabular}
\end{center}
\end{table}

Table \ref{tab:int_fluxes}
presents the integrated line fluxes for these apertures, 
as $\int \!\!\! \int T_{B} \, \mathrm{d}v \, \mathrm{d}A$ in 
K~km~s$^{-1}$~pc$^{2}$. 
The brightness temperature $T_{B}$ is the extended beam temperature,
corrected for the extended aperture efficiency of $\sim$~0.65 \citep{ur+10},
as discussed below. 

The line emission is integrated over velocity, with
the accuracy limited by low level baseline ripples in the spectra, 
as discussed in subsection \ref{subsec:cubes}. To reduce this problem,
we integrated over a velocity range which covers the expected emission 
while excluding too much extra data in the cubes. Hence this velocity range
varies with the area (due to differential velocity across the CMZ, and different
velocity widths in the different areas) and the line. Uncertainties are quoted,
based on observed integrated intensities fluctuations, caused by these baseline 
offsets, over parts of the data cubes without lines. However, we do note that
the baseline ripples vary between the different lines, as does the noise
in the cubes (Table \ref{tab:summary_table}), so that these uncertainty
estimates are only indicative.

These line luminosities can be compared to those
of 3-mm lines over the same areas in \citet{jo+12}, and are quoted 
in the form of $\int \!\!\! \int T_{B} \, \mathrm{d}v \, \mathrm{d}A$
for ease of comparison with other galaxies. 

We do not include all the lines
in this table, but exclude the weakest lines which are generally below
the 3~$\sigma$ level of significance in the areas given the baseline problems 
(H$^{13}$CCCN, HCC$^{13}$CN, HC$_{5}$N, SiO masers, RRLs).

Table \ref{tab:rel_fluxes} presents the integrated line fluxes over the
six apertures, normalised by the value for the CS line.
The Sgr B2 area generally has an enhanced line flux ratio to CS, as compared
to the other areas, and the sub-area Sgr B2 Core has even more enhanced line 
ratios. While we expect Sgr B2 to be chemically rich, the high ratios relative
to CS are 
likely to be in part due to the high optical depth of CS in this area.
This can be checked with the isotopologues $^{13}$CS and C$^{34}$S.
These do indeed show higher C$^{34}$S/CS ratio in Table \ref{tab:rel_fluxes}
at Sgr B2, and particularly at 
the Core, which indicates high optical depth in CS. Expressed in the more commonly
used form of the ratio CS/C$^{34}$S (i.e. the inverse to that tabulated) 
this is lower ratio CS/C$^{34}$S in Sgr B2 (8.8) and the Sgr B2 Core (7.5)
compared to the other regions. 
The situation in the weaker
$^{13}$CS line is less clear, with lower CS/$^{13}$CS at the Sgr B2 Core,
of 14 compared to the expected optically thin value of 24
for the Galactic Centre from \citet{lape90}, but overall ratio 20 in the 
larger Sgr B2 region (and an anomalously high ratio 63 for Sgr C, which may
indicate baselevel problems).

\begin{table*}
\begin{center}
\caption{Integrated line fluxes as 
$\int \!\!\! \int T_{B} \, \mathrm{d}v \, \mathrm{d}A$, integrated over 
velocity, and 
the areas given in Table \ref{tab:apertures}. The integrated fluxes are in units
10$^{3}$~K~km~s$^{-1}$~pc$^{2}$, The 1~$\sigma$
uncertainties are given, and the upper limits are at the 3~$\sigma$ level.}
\label{tab:int_fluxes}
\begin{tabular}{ccccccc}
\hline
Line        & CMZ            & Sgr A          & Sgr B2         & Sgr C           
& G1.3           & Sgr B2 Core \\
\hline
CS          & 1590 $\pm$  80 &  218 $\pm$  11 & 297  $\pm$ 15  &  157 $\pm$ 8  
&  246 $\pm$ 12  & 33.2 $\pm$ 1.7 \\ 
CH$_{3}$OH  &  650 $\pm$  30 &   66 $\pm$   4 & 182  $\pm$  9  &   48 $\pm$ 2.5  
&  130 $\pm$ 7   & 29.4 $\pm$ 1.5 \\
HC$_{3}$N   &  287 $\pm$  15 & 33.2 $\pm$ 2.2 &  82  $\pm$  5  & 16.6 $\pm$ 1.0 
&   57 $\pm$ 3   & 15.1 $\pm$ 0.8 \\
SiO         &  283 $\pm$  15 & 24.8 $\pm$ 1.9 &  71  $\pm$  4  & 14.4 $\pm$ 1.0 
&   72 $\pm$ 4   &  8.5 $\pm$ 0.5 \\
HNCO        &  312 $\pm$  16 & 23.2 $\pm$ 1.8 &  80  $\pm$  5  & 14.6 $\pm$ 1.0 
&   66 $\pm$ 3   & 12.9 $\pm$ 0.7 \\
HOCO$^{+}$  &   30 $\pm$   5 &  $<$ 4         &  8.0 $\pm$ 2.5 & $<$ 1.9        
&  5.5 $\pm$ 0.9 & 1.69 $\pm$ 0.28 \\
NH$_{2}$CHO &   20 $\pm$   5 &  $<$ 4         &  7.9 $\pm$ 2.5 & $<$ 1.9        
&  4.1 $\pm$ 0.9 & 1.36 $\pm$ 0.28 \\
OCS         &   60 $\pm$   6 &  $<$ 4         & 25.2 $\pm$ 2.7 & $<$ 1.9        
& 16.6 $\pm$ 1.2  &  3.6 $\pm$ 0.3 \\
HCS$^{+}$   &   27 $\pm$   5 &  5.2 $\pm$ 1.4 &  $<$ 7         & $<$ 1.9        
&  3.9 $\pm$ 0.9  & 1.11 $\pm$ 0.28 \\
CCS         &   34 $\pm$   6 &  $<$ 4         & 13.5 $\pm$ 2.5 & $<$ 1.9        
&  8.4 $\pm$ 0.9  & 2.00 $\pm$ 0.29 \\
C$^{34}$S   &  114 $\pm$   8 &  9.7 $\pm$ 1.5 & 33.5 $\pm$ 2.9 & 10.5 $\pm$ 0.8 
& 19.8 $\pm$ 1.3  &  4.4 $\pm$ 0.4 \\
$^{13}$CS   &   56 $\pm$   6 & 11.6 $\pm$ 1.5 & 14.8 $\pm$ 2.5 &  2.5 $\pm$ 0.6 
&  9.3 $\pm$ 1.0  &  2.4 $\pm$ 0.3 \\
$^{29}$SiO  &    22 $\pm$  5 &  $<$ 4         &  $<$ 7         &  2.0 $\pm$ 0.6 
&  3.0 $\pm$ 0.9 &  $<$ 0.8 \\
CH$_{3}$OH maser & 53 $\pm$ 6 & 9.6 $\pm$ 1,5 & 12.0 $\pm$ 2.5 &  4.0 $\pm$ 0.6
& 10.1 $\pm$ 1.0 &  4.0 $\pm$ 0.3 \\
\hline
\end{tabular}
\end{center}
\end{table*}

\begin{table*}
\begin{center}
\caption{Integrated line fluxes, integrated over velocity and 
the areas given in Table \ref{tab:apertures}, normalised to CS. The 1~$\sigma$
uncertainties are given, and the upper limits are at the 3~$\sigma$ level.}
\label{tab:rel_fluxes}
\begin{tabular}{ccccccc}
\hline
Line         & CMZ               & Sgr A             & Sgr B2            & 
Sgr C            & G1.3         & Sgr B2 Core \\
\hline
CS           &  1.00 def.        &  1.00 def.        & 1.00 def.         &
1.00 def.         & 1.00 def.         & 1.00 def.          \\
CH$_{3}$OH   &  0.41 $\pm$ 0.03  & 0.301 $\pm$ 0.022 &  0.61 $\pm$  0.05 &
0.306 $\pm$ 0.022 &  0.53 $\pm$ 0.04  &  0.89 $\pm$ 0.06 \\
HC$_{3}$N    & 0.181 $\pm$ 0.013 & 0.152 $\pm$ 0.013 & 0.278 $\pm$ 0.021 &
0.106 $\pm$ 0.008 & 0.230 $\pm$ 0.017 &  0.46 $\pm$ 0.03 \\
SiO          & 0.178 $\pm$ 0.013 & 0.113 $\pm$ 0.010 & 0.238 $\pm$ 0.019 &
0.092 $\pm$ 0.008 & 0.291 $\pm$ 0.021 & 0.255 $\pm$ 0.020 \\
HNCO         & 0.197 $\pm$ 0.014 & 0.106 $\pm$ 0.010 & 0.270 $\pm$ 0.021 &
0.093 $\pm$ 0.008 & 0.270 $\pm$ 0.019 &  0.39 $\pm$  0.03 \\
HOCO$^{+}$   & 0.019 $\pm$ 0.004 &  $<$ 0.019        & 0.027 $\pm$ 0.008 &
  $<$ 0.012       & 0.022 $\pm$ 0.004 & 0.051 $\pm$ 0.009 \\
NH$_{2}$CHO  & 0.013 $\pm$ 0.003 &  $<$ 0.019        & 0.027 $\pm$ 0.008 &
  $<$ 0.012       & 0.017 $\pm$ 0.004 & 0.041 $\pm$ 0.009 \\
OCS          & 0.038 $\pm$ 0.004 &  $<$ 0.019        & 0.085 $\pm$ 0.010 &
  $<$ 0.012       & 0.067 $\pm$ 0.006 & 0.109 $\pm$ 0.011 \\
HCS$^{+}$    & 0.017 $\pm$ 0.004 & 0.024 $\pm$ 0.007 &  $<$ 0.024        &
  $<$ 0.012       & 0.016 $\pm$ 0.004 & 0.033 $\pm$ 0.008 \\
CCS          & 0.021 $\pm$ 0.004 &  $<$ 0.019        & 0.046 $\pm$ 0.009 &
  $<$ 0.012       & 0.034 $\pm$ 0.004 & 0.060 $\pm$ 0.009 \\
C$^{34}$S    & 0.072 $\pm$ 0.006 & 0.044 $\pm$ 0.007 & 0.113 $\pm$ 0.011 &
0.067 $\pm$ 0.006 & 0.080 $\pm$ 0.007 & 0.133 $\pm$ 0.012 \\
$^{13}$CS    & 0.035 $\pm$ 0.004 & 0.053 $\pm$ 0.007 & 0.050 $\pm$ 0.009 &
0.016 $\pm$ 0.004 & 0.038 $\pm$ 0.004 & 0.073 $\pm$ 0.010 \\
$^{29}$SiO   & 0.014 $\pm$ 0.003 &  $<$ 0.019        &  $<$ 0.024        &
0.013 $\pm$ 0.004 & 0.012 $\pm$ 0.003 & $<$ 0.024         \\
CH$_{3}$OH maser & 0.033 $\pm$ 0.004 & 0.044 $\pm$ 0.007 & 0.041 $\pm$ 0.009 &
0.025 $\pm$ 0.004 & 0.041 $\pm$ 0.004 & 0.121 $\pm$ 0.012 \\
\hline
\end{tabular}
\end{center}
\end{table*}

\subsection{Ratio of 7-mm and 3-mm lines}
\label{subsec:ratio}

The observed integrated line intensity $\int T \mathrm{d}v$ of a spectral line
can be related to the column density $N_{u}$ of molecules in the upper level 
of the transition, for optically thin lines
using standard assumptions of LTE radiative transfer \citep*{wirohu09}, by
$ N_{u} = (8 \pi \nu^2 k/h c^3 A_{ul}) \int T_{B} \mathrm{d}v $, 
where $A_{ul}$ is the Einstein coefficient. 
The total column density $N$ of the molecule is given by
$ N = (N_{u}/g_{u}) Q_T \exp(E_{u}/kT_{ex}) $
where $Q_{T}(T_{ex})$ is the partition function, $T_{ex}$ 
is the excitation temperature, 
$E_{u}$ is the energy of the upper level and $g_{u}$ is the statistical weight of 
the upper level.

The excitation temperature $T_{ex}$ describes the relative populations
in different levels, and is typically not the same as the gas kinetic 
temperature $T_{kin}$, as at low densities the excitation depends on
the radiation as well as collisions \citep{wirohu09}.

As we have 3-mm line observations of the CMZ from \citet{jo+12} to match the
7-mm observations here, we can consider the ratio of integrated line 
intensities $\int T \mathrm{d}v$, which is related (subject to assumptions, and 
caveats below) to the excitation temperature $T_{ex}$. 

We consider five molecules, that is HNCO, HOCO$^{+}$, HC$_{3}$N, SiO and 
$^{13}$CS, common to both our 7-mm and 3-mm CMZ spectral imaging.
The 3-mm data were smoothed to the resolution of the 7-mm data, using the
beamsizes (62 to 68 arcsec) interpolated from the Mopra 7-mm measurements 
of \citet{ur+10}.
The 3-mm and 7-mm data expressed as $T_{A}^{*}$ were converted to
brightness temperatures $T_{B}$ appropriate for extended emission which
fills both the main beams and inner sidelobes, by dividing by the extended
beam efficiencies at 3-mm (0.62 to 0.65) and 7-mm (0.66 to 0.69)
interpolated from \citet{la+05} and \citet{ur+10} respectively.

The ratio image of $ \int T_B \mathrm{d}v $ for the 3-mm and 7-mm lines was 
taken by summing 3-dimensional pixels (voxels) over velocity in both 
cubes, over the region where both lines 
were above the respective 3-$\sigma$ level, and then dividing the integrated 
emission images. This reduces the effect of including data without 
significant emission which adds artifacts due to low-level baseline offsets
(particularly in the 3-mm data which is more weather affected), but should
not bias the ratio due to missing flux, as we consider the ratio of emission 
from the same voxels. It is effectively a weighted mean of the ratio, considered
on a voxel basis, with the weighting set to zero below the 3-$\sigma$ level.

\begin{figure*}
\includegraphics[angle=-90,width=16.8cm]{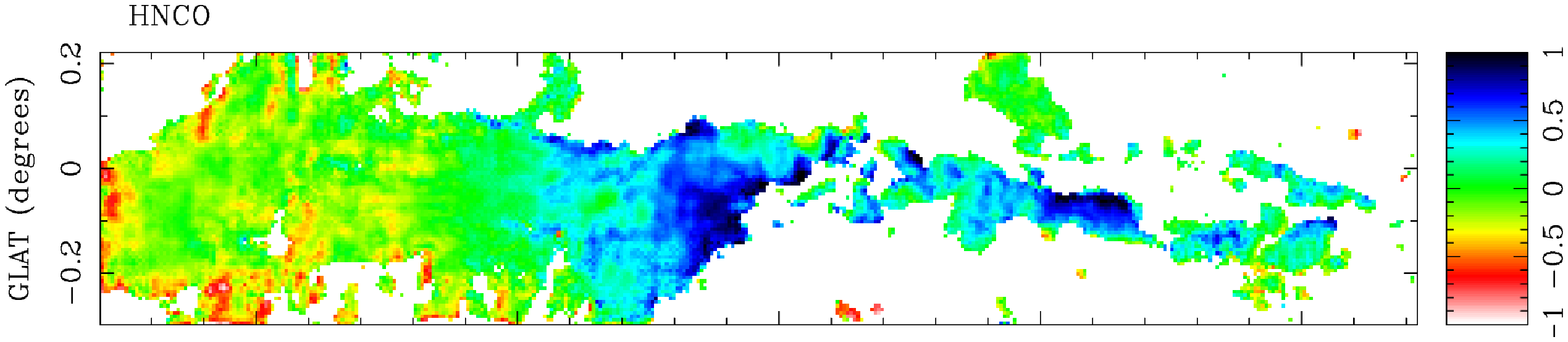}
\includegraphics[angle=-90,width=16.8cm]{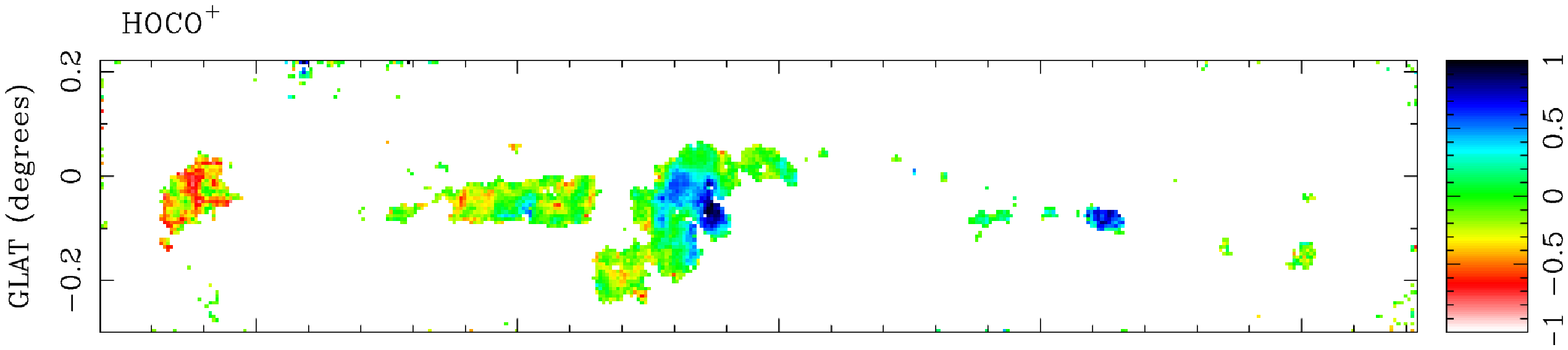}
\includegraphics[angle=-90,width=16.8cm]{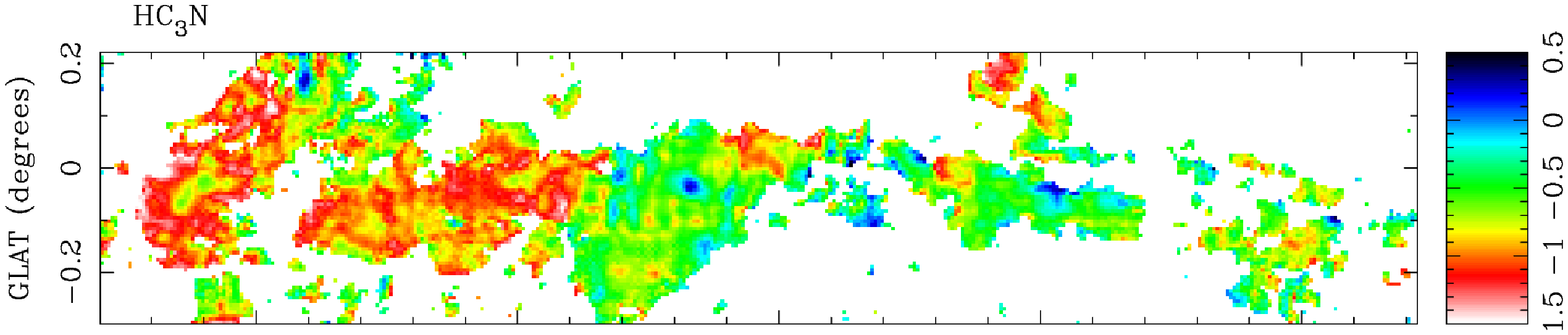}
\includegraphics[angle=-90,width=16.8cm]{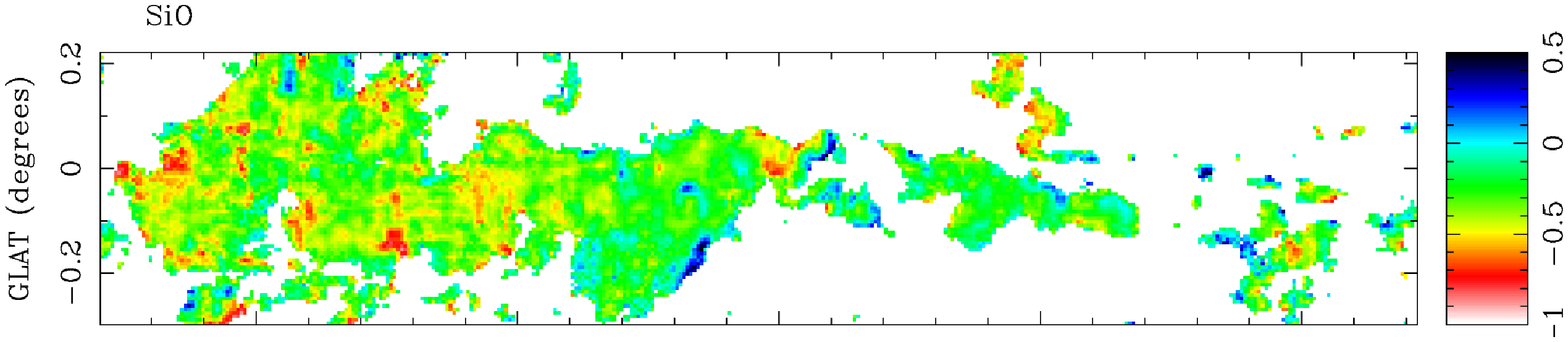}
\includegraphics[angle=-90,width=16.8cm]{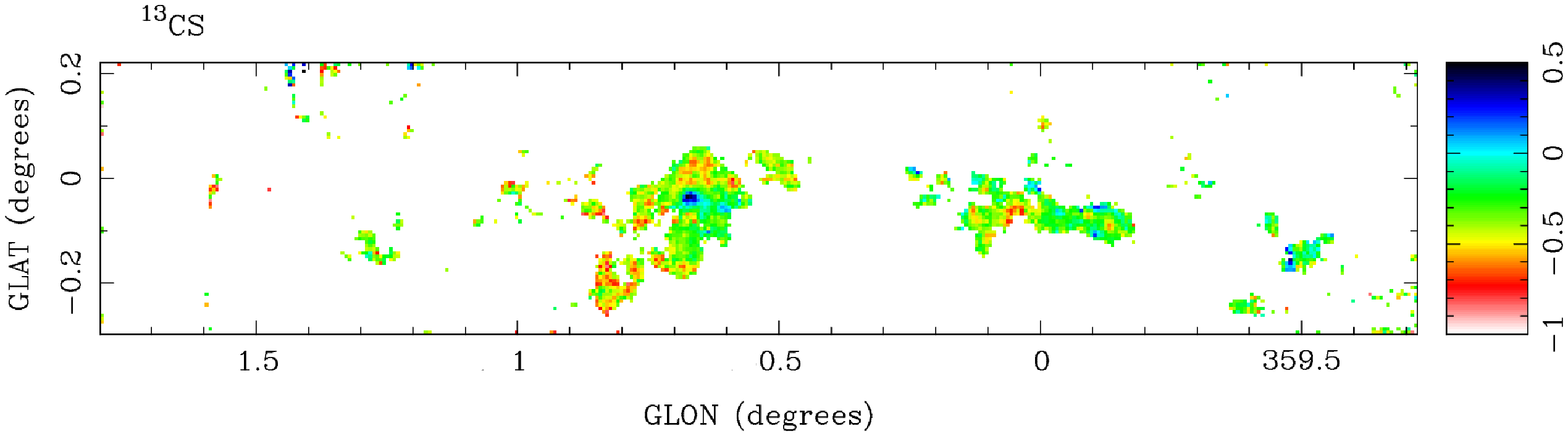}
\caption{The distribution of the ratio of integrated line brightness of 3-mm 
to 7-mm lines of HNCO, HOCO$^{+}$, HC$_{3}$N, SiO and $^{13}$CS expressed
as $\ln (\int T_{B, 3mm} \mathrm{d}v / \int T_{B, 7mm} \mathrm{d}v)$.}
\label{fig:ratios}
\end{figure*}

The ratio images are presented in Fig. \ref{fig:ratios} as the natural 
logarithm of the ratio of 3-mm to 7-mm $ \int T_{B} \mathrm{d}v $, 
$\ln (\int T_{B, 3mm} \mathrm{d}v / \int T_{B, 7mm} \mathrm{d}v)$. The HNCO, HOCO$^{+}$ and
HC$_{3}$N images show variations in the ratio across the CMZ, in the sense
that the 3-mm to 7-mm ratio is higher around Sgr~B2 and Sgr~A, and is less
around longitudes 1.0 deg. to 1.8 deg. There are some signs of similar variation
in the ratio in the SiO image, but less clearly.

\begin{figure*}
\includegraphics[angle=-90,width=8.0cm]{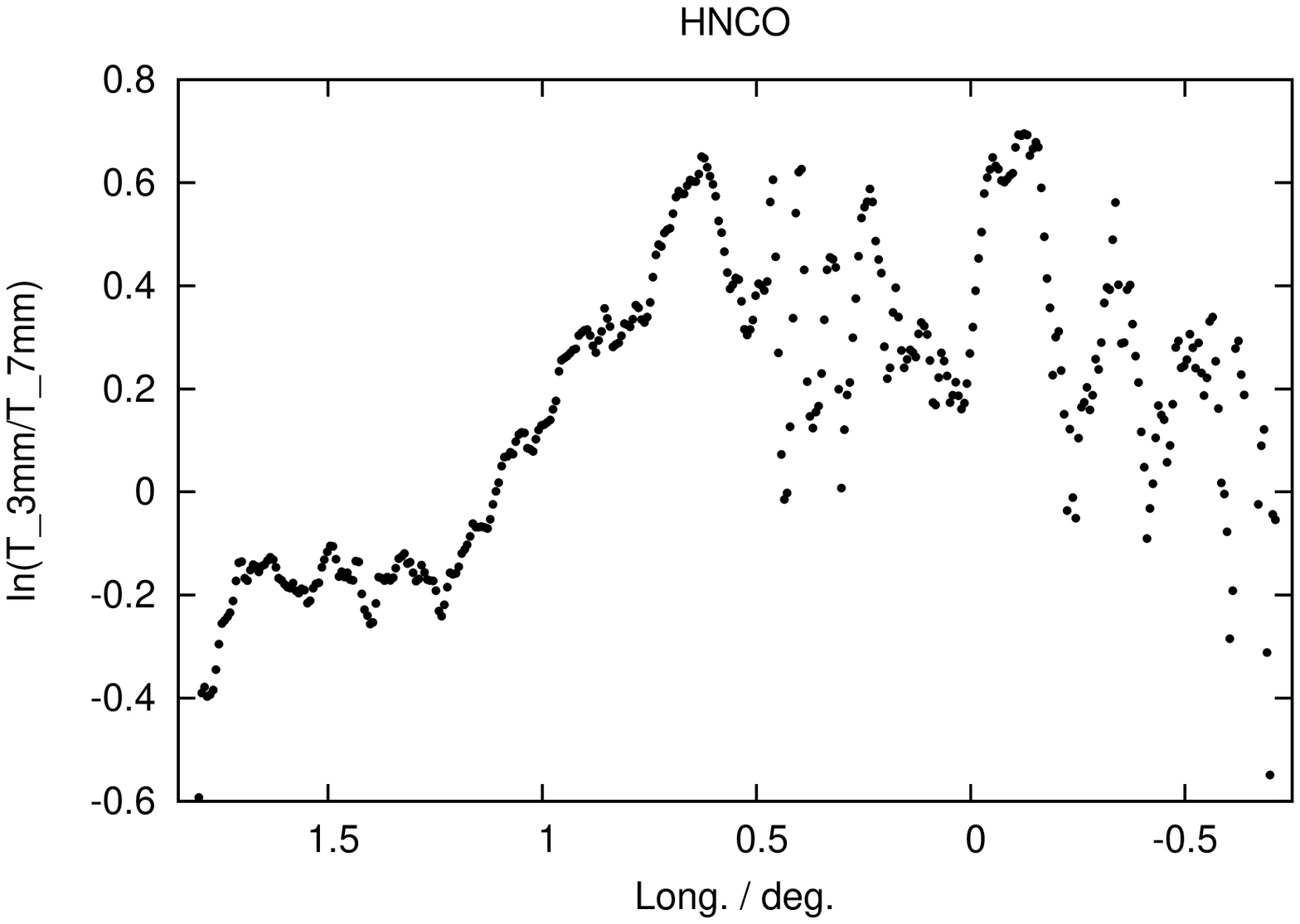}
\includegraphics[angle=-90,width=8.0cm]{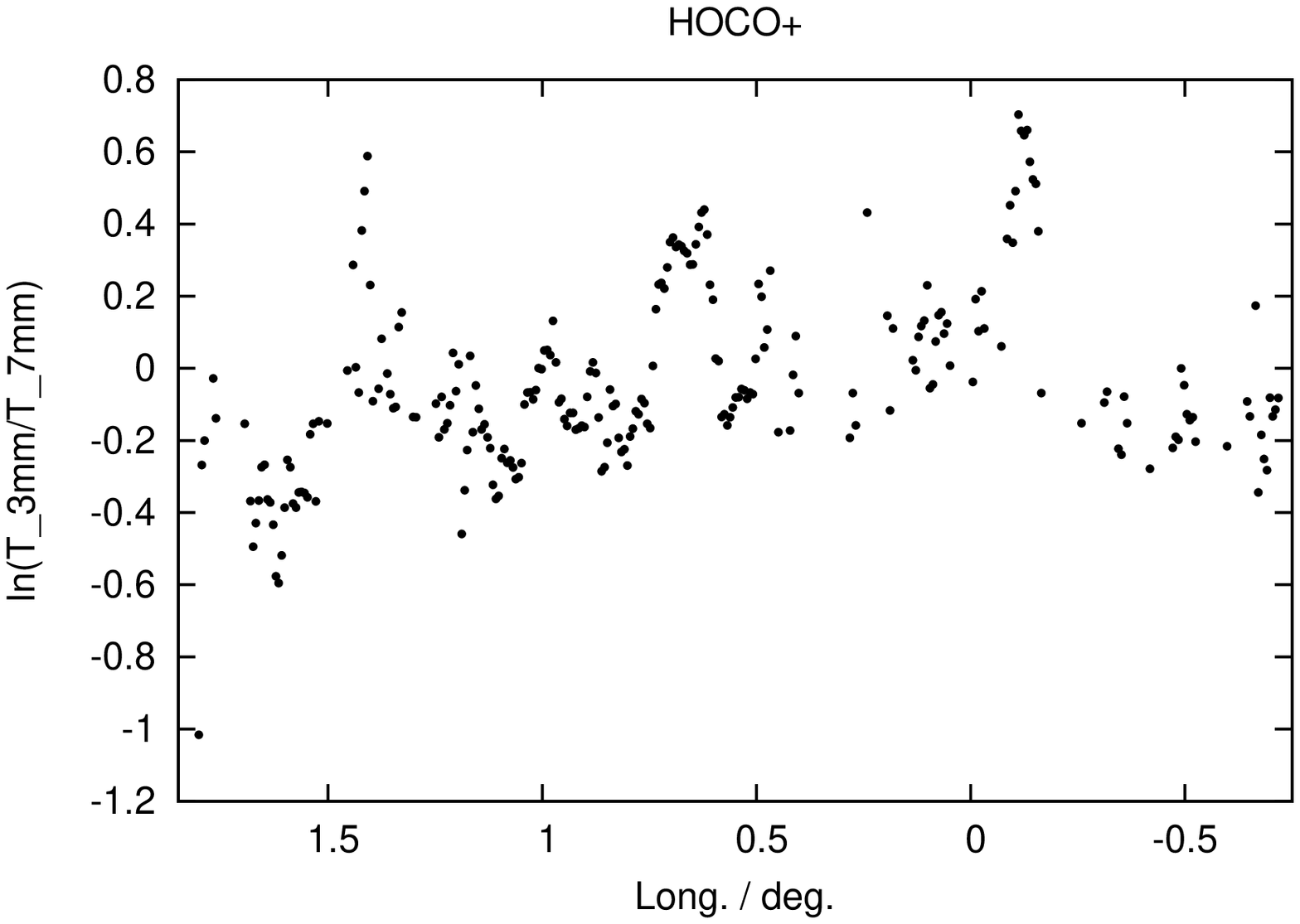}
\includegraphics[angle=-90,width=8.0cm]{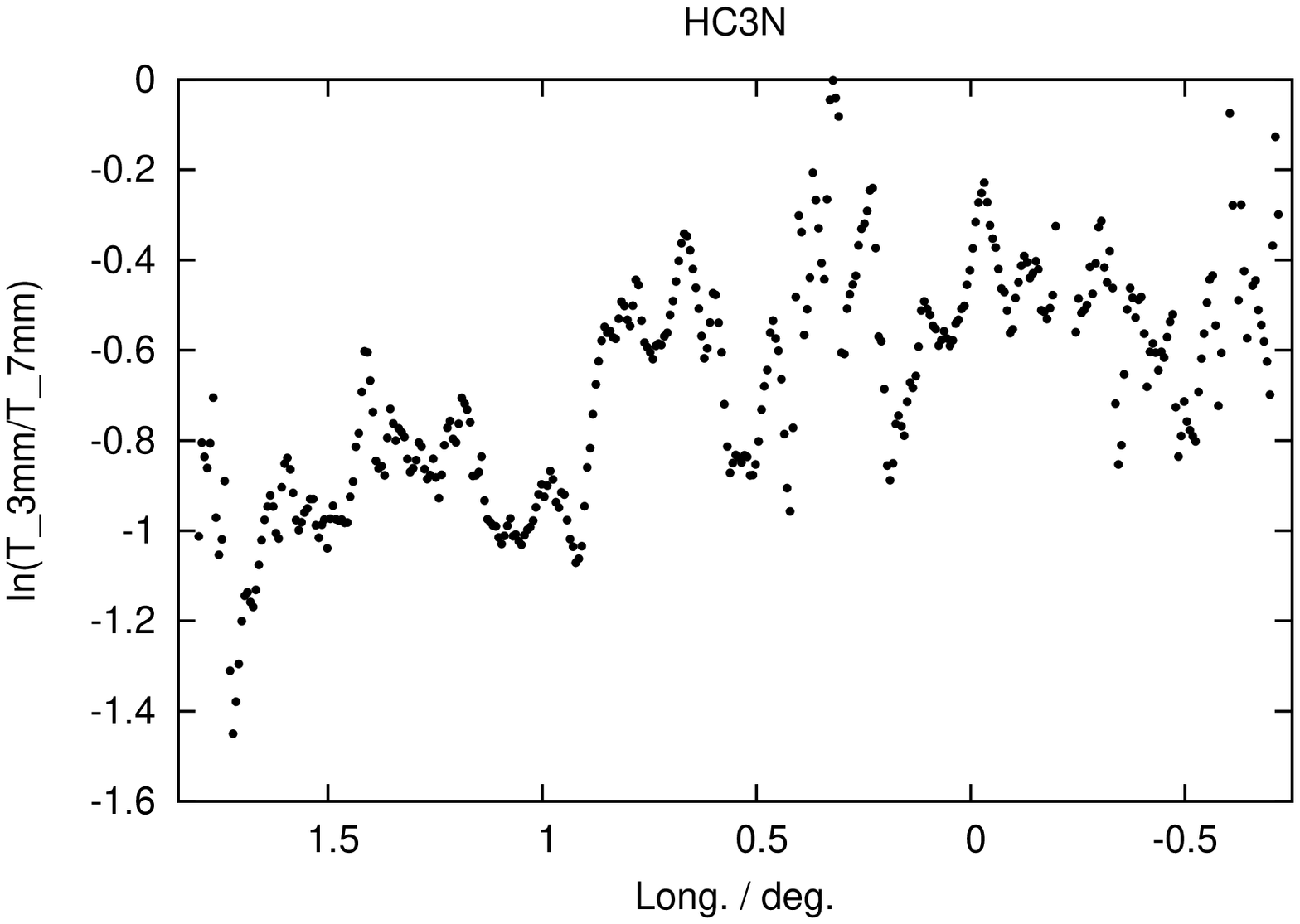}
\includegraphics[angle=-90,width=8.0cm]{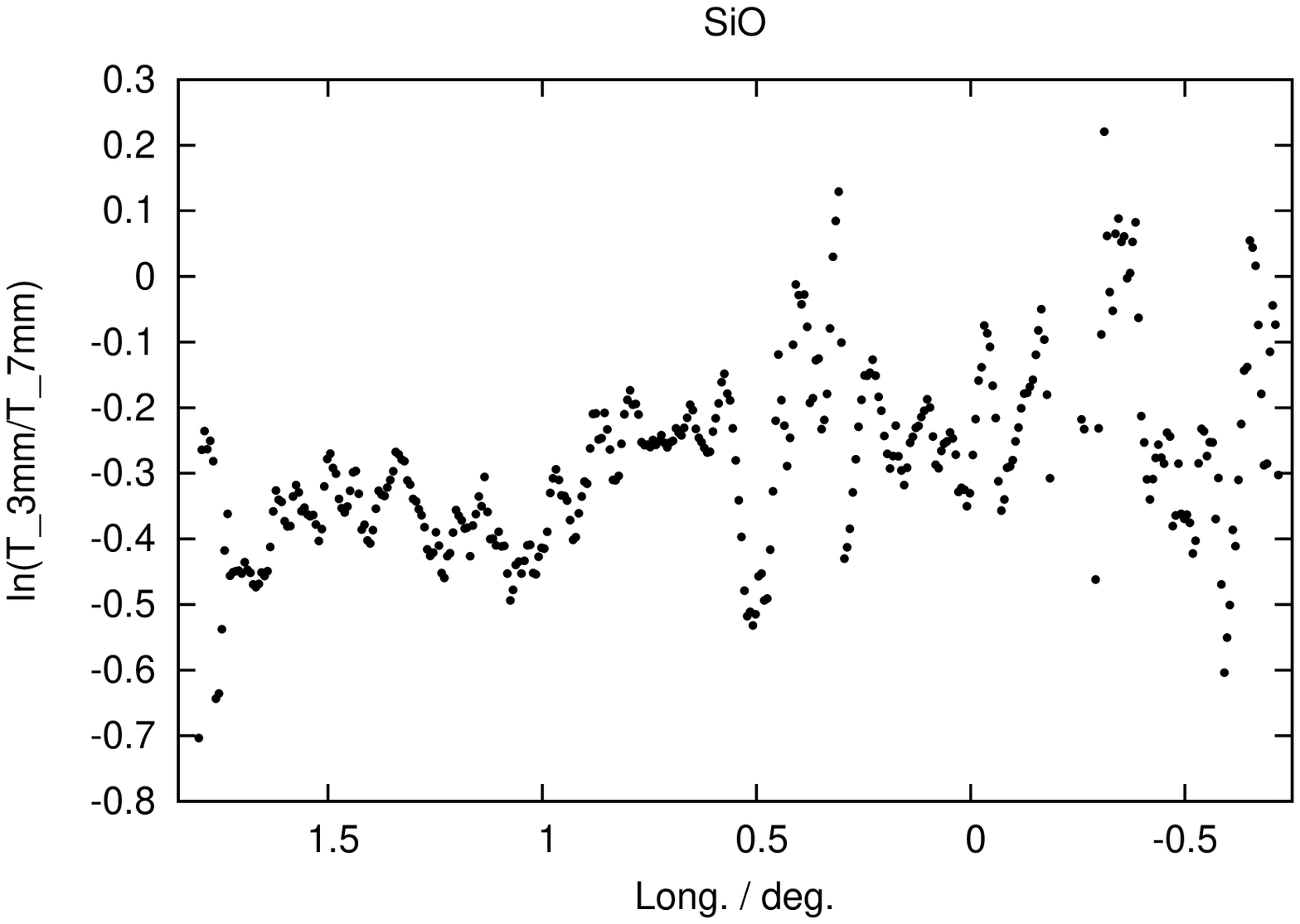}
\includegraphics[angle=-90,width=8.0cm]{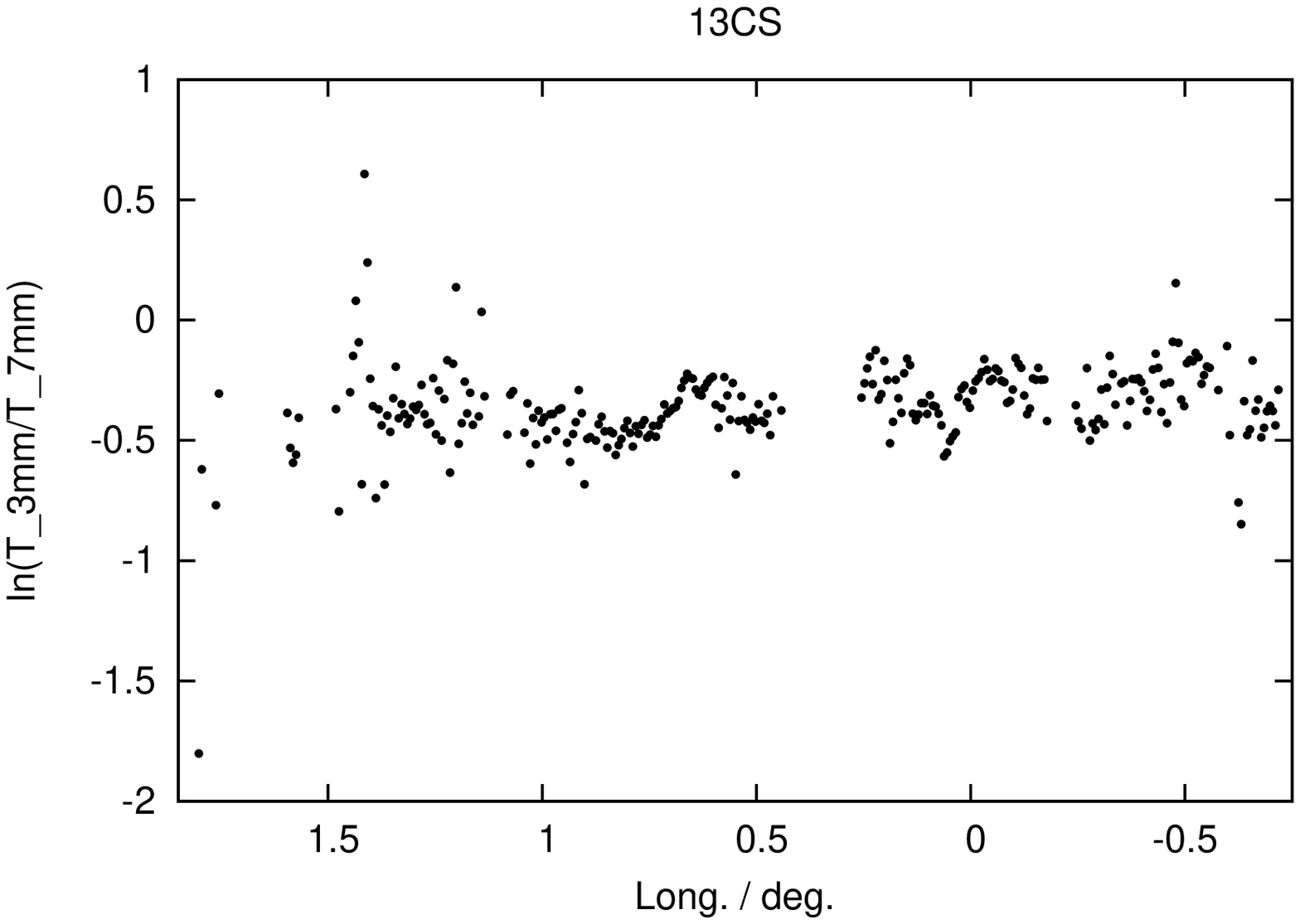}
\caption{The ratio of integrated line brightness of 3-mm 
to 7-mm lines of HNCO, HOCO$^{+}$, HC$_{3}$N, SiO and $^{13}$CS expressed
as $\ln (\int T_{B, 3mm} \mathrm{d}v / \int T_{B, 7mm} \mathrm{d}v)$,
averaged over latitude, to show the variation with longitude. 
Note the variation in line ratio across the CMZ, most clearly shown in 
HNCO, HC$_{3}$N and SiO.}
\label{fig:ratio_aver}
\end{figure*}

The ratios are plotted in Fig. \ref{fig:ratio_aver} averaged over latitude,
to show the longitude variations more clearly. The plots for HOCO$^{+}$ and $^{13}$CS
are noisy for longitude ranges with little data, but the HOCO$^{+}$ does show
the difference in ratio between Sgr~B2 and higher longitudes. The plots for
HNCO, HC$_{3}$N and SiO do show ratio variations.

At the centre of Sgr~B2 in Fig. \ref{fig:ratios} there is a small region of 
higher 3-mm to 7-mm ratio
in HC$_{3}$N, SiO and $^{13}$CS, associated with the free-free radio peak
of Sgr~B2. We attribute this to the radiative transfer effect of line
emission being absorbed in front of the strong continuum. This occurs
to a greater extent in the 7-mm band compared to the 3-mm band, as discussed in
\citet{jo+12}, leading to the enhanced  3-mm to 7-mm ratio.

The line ratio differences can largely be interpreted as excitation
temperature differences, as discussed above, with the caveat that other
effects may be occuring, such as the continuum absorption and signifcant 
optical depth (subsection \ref{subsec:PCA}). The line frequencies ($\nu$), 
Einstein coefficients ($A_{ul}$), statistical weights ($g_{u}$) and upper
energy levels ($E_{u}$) were obtained from the splatalogue online compilation
(http://www.splatalogue.net/). 

The statistics of the line ratios 
$\ln (\int T_{B, 3mm} \mathrm{d}v / \int T_{B, 7mm} \mathrm{d}v)$ 
(mean and standard deviation) are listed in Table \ref{tab:ratios}, for the areas
in Table \ref{tab:apertures}and Fig. \ref{fig:areas}. 
The corresponding excitation
temperatures $T_{ex}$ assuming optically thin LTE
conditions are listed in Table \ref{tab:Tex}.

Note that these calculated excitation temperatures $T_{ex}$ are much lower
than the kinetic temperature $T_{kin} \sim$~ 30~K. In practice,
for the likely physical conditions in the CMZ, the energy
levels of these molecules are unlikely to be in LTE (see below, 
section \ref{sec:disc}). Hence the values of 
$T_{ex}$ calculated here
are a useful way to describe the relative populations of the levels
that give rise to the 7-mm and 3-mm lines, but should not be over-interpreted
as applying to other levels.

We note the line ratios are smaller beyond Sgr~B2, and thus outside 
the  $\rm 10^7\,M_{\odot}$
rotating dust ring, that extends from Sgr B2 to Sgr A, citep{mo+11}.
This is particularly noticeable
with the G1.3 dust core, where the higher excitation 3-mm lines are
relatively weaker.

\begin{table*}
\begin{center}
\caption{Mean and standard deviation of the ratio of 3-mm to 7-mm lines,
expressed as $\ln$(ratio), ie $\ln (T_{3mm} / T_{7mm})$ in the selected areas.}
\label{tab:ratios}
\begin{tabular}{ccccccc}
\hline
Molecule  & CMZ  & Sgr A  & Sgr B2 & 
Sgr C  & G1.3  & Sgr B2 Core \\
\hline
HNCO       & ~~$ 0.07 \pm 0.32$ & ~~$ 0.37 \pm 0.24 $ & ~~$ 0.43 \pm 0.21 $ & 
~~$ 0.16 \pm 0.25 $ ~~& $ -0.14 \pm 0.13 $ ~~ & ~~$ 0.58 \pm 0.11 $ \\
HOCO$^{+}$ & $-0.04 \pm 0.33$ & ~~$ 0.28 \pm 0.29 $ & ~~$ 0.07 \pm 0.29 $ & 
$ -0.12 \pm 0.15 $ \# & $ -0.13 \pm 0.14 $ \# & ~~$ 0.39 \pm 0.23 $ \\
HC$_{3}$N  & $-0.71 \pm 0.32$ & $ -0.48 \pm 0.24 $ & $ -0.63 \pm 0.28 $ & 
$ -0.58 \pm 0.25 $ ~~& $ -0.83 \pm 0.28 $ ~~ & $ -0.45 \pm 0.23 $ \\
SiO        & $-0.30 \pm 0.18$ & $ -0.22 \pm 0.14 $ & $ -0.25 \pm 0.15 $ & 
$ -0.21 \pm 0.21 $ ~~& $ -0.34 \pm 0.14 $ ~~ & $ -0.25 \pm 0.13 $ \\
$^{13}$CS  & $-0.36 \pm 0.25$ & $ -0.30 \pm 0.16 $ & $ -0.37 \pm 0.17 $ & 
$ -0.23 \pm 0.18 $ ~~& $ -0.35 \pm 0.16 $ \# & $ -0.31 \pm 0.22 $ \\
\hline
\end{tabular}

\# = small number of pixels ($< 100$) with good data (clipped at 3~$\sigma$.)
\end{center}
\end{table*}

\begin{table*}
\begin{center}
\caption{The excitation
temperatures $T_{ex}$ in K, assuming optically-thin LTE conditions, derived from the
ratio of 3-mm to 7-mm lines. Note, however,
that {\sc radex} modelling in section \ref{sec:disc} shows that 
assuming optically-thin LTE is too simplistic.}
\label{tab:Tex}
\begin{tabular}{ccccccc}
\hline
Molecule         & CMZ               & Sgr A             & Sgr B2            & 
Sgr C            & G1.3         & Sgr B2 Core \\
\hline
HNCO       & $5.6^{+1.8}_{-1.1}$ & $7.3^{+2.3}_{-1.4}$ & $7.7^{+2.2}_{-1.4}$ &
 $6.0^{+1.6}_{-1.0}$ ~~ & $4.8^{+0.5}_{-0.4}$ ~~ & $9.2^{+1.5}_{-1.1}$ \\
HOCO$^{+}$ & $5.0^{+1.5}_{-1.0}$ & $6.5^{+2.3}_{-1.4}$ & $5.5^{+1.5}_{-1.0}$ &
 $4.8^{0.5+}_{-0.4}$ \# & $4.7^{+0.5}_{-0.4}$ \# & $7.2^{+2.2}_{-1.4}$ \\
HC$_{3}$N  & $8.9^{+1.3}_{-1.0}$ & $9.4^{+1.4}_{-1.1}$ & $8.7^{+1.4}_{-1.1}$ &
 $8.9^{+1.3}_{-1.0}$ ~~ & $7.9^{+1.1}_{-0.9}$ ~~ & $9.5^{+1.4}_{-1.1}$ \\
SiO        & $2.5^{+0.3}_{-0.2}$ & $2.6^{+0.3}_{-0.2}$ & $2.6^{+0.3}_{-0.2}$ &
 $2.6^{+0.4}_{-0.3}$ ~~ & $2.4^{+0.2}_{-0.2}$ ~~ & $2.6^{+0.2}_{-0.2}$ \\
$^{13}$CS  & $2.2^{+0.4}_{-0.3}$ & $2.2^{+0.2}_{-0.2}$ & $2.2^{+0.2}_{-0.2}$ &
 $2.3^{+0.3}_{-0.2}$ ~~ & $2.2^{+0.4}_{-0.3}$ \# & $2.2^{+0.3}_{-0.3}$ \\
\hline
\end{tabular}

\# = small number of pixels ($< 100$) with good data (clipped at 3~$\sigma$.)
\end{center}
\end{table*}

\section{Discussion}
\label{sec:disc}

The kinetic temperature $T_{kin}$ and hydrogen density $n$ in the CMZ
have been fitted by \citet{na+07} using CO lines ($^{12}$CO 1--0, $^{12}$CO
3--2, $^{13}$CO 1--0) with a Large Velocity Gradient (LVG) model, giving
typical values $T_{kin} = 30$~K and $n = 3 \times 10^{3}$~cm$^{-3}$.

For such physical conditions $T_{kin}$ and $n$ we can confirm with the
non-LTE model {\sc RADEX} of \citet{va+07}, that the 7-mm and 3-mm lines studied 
in section \ref{subsec:ratio} give a low effective excitation temperature
$T_{ex}$. The hydrogen density $n$ is much lower than the critical
density $n_{crit}$ at which the collisions would thermalise the molecular
levels to $T_{kin}$. We used the online version of 
{\sc RADEX}\footnote{http://www.sron.rug.nl/\~vdtak/radex/radex.php}, to explore
the parameter space that would reproduce the 7-mm line fluxes and the 
3-mm to 7-mm line ratios. We do note that this still assumes an isothermal 
homogeneous medium. 

The input parameters to {\sc RADEX} are excitation
conditions, $T_{kin}$, $n$ and background temperature $T_{bg}$, and
radiative transfer parameters, $N$ and line width $\Delta\,v$. If we fix
$T_{kin} = 30$~K, $\Delta\,v = 10$~km\,s$^{-1}$ which is typical of these
lines in the CMZ, and $T_{bg} = 2.73$~K for the microwave background
temperature, then we have two free parameters $n$ and $N$. Varying $n$
changes the relative populations of the molecular levels, and hence
the ratio of 3-mm to 7-mm transitions. Varying $N$ changes the line
brightness temperature $T_{R}$. 
With two free parameters, we can fit the
two observable parameters, but only if the combination of observable
parameters is in the range of the model, so this is not a trivial process.
In practice, varying one of $n$ and $N$
changes both the ratio and $T_{R}$ due to non-linearities, such as optical
depth.

The results of such modelling is given in Table \ref{tab:radex} for
four molecules. HOCO$^{+}$ is not one of the molecules with data files
available for {\sc RADEX}. We varied $n$ and $N$ to obtain typical values of 
$\ln (T_{B, 3mm} / T_{B, 7mm})$ and $T_{R}$ (for the 7-mm line) given in the 
data in subsection \ref{subsec:ratio}. For each molecule, we present four fits,
giving a typical $T_{R}$ with mean ratio, higher and lower ratios by 0.3
in the $\ln$(ratio) or factor 1.35, and a fit for mean ratio and higher $T_{R}$.

The derived values of $n$ are a few times 
10$^{4}$ cm$^{-3}$, which is somewhat larger than the typical value
$n = 3 \times 10^{3}$~cm$^{-3}$ obtained by \citet{na+07} for CO, but
these molecules with higher $n_{crit}$ preferentially trace higher demsities. 

We have shown here that the observed ratio of the 3-mm to 7-mm lines, and the
line brightness temperatures, can be fitted with plausible values of $n$ and $N$.
With only two lines, we can only constrain two parameters at a time, with
other RADEX input parameters fixed. In practice, it is quite likely
that the kinetic temperature  $T_{kin}$ and background temperature $T_{bg}$
vary across the CMZ as well, so the actual variation in physical conditions
is more complcated.

The higher 3-mm to 7-mm line ratios around Sgr~B2 and Sgr~A, compared to
longitudes around G1.3, can be partly explained by higher density $n$,
which means that the sub-thermal excitation gets closer to 
thermalisation with kinetic temperature $T_{kin}$.
As noted above, however, from the data and analysis here we cannot rule out
changes in $T_{bg}$ and $T_{kin}$ too.


\begin{table}
\begin{center}
\caption{Results of {\sc RADEX} non-LTE models to reproduce typical 3-mm to 7-mm
line ratios, and 7-mm line brightness observed in the CMZ. These are labelled
m for mean or mid-range, l for low and h for high. We assume fixed
$T_{kin} = 30$~K, $\Delta\,v = 10$~km\,s$^{-1}$ and $T_{bg} = 2.7$~K.}
\label{tab:radex}
\begin{tabular}{ccccc}
\hline
Molecule & $n$  & $N$ & $\ln (T_{3mm} / T_{7mm})$ & $T_{R}$ \\
         & 10$^{4}$ cm$^{-3}$ & 10$^{14}$ cm$^{-2}$ &      &  K   \\
\hline
HNCO      &  0.65 &  1.7  & $-0.24$ ~l &  1.0  ~m \\
HNCO      &  1.1  &  1.7  & ~~$0.08$ ~m &  1.0  ~m \\
HNCO      &  1.7  &  1.8  & ~~$0.37$ ~h &  1.0  ~m \\
HNCO      &  1.2  &  4.5  & ~~$0.08$ ~m &  2.5  ~h \\
          &       &       &            &          \\
HC$_{3}$N &  1.8  &  0.50 & $-1.02$ ~l &  1.0  ~m \\
HC$_{3}$N &  2.7  &  0.50 & $-0.70$ ~m &  1.0  ~m \\
HC$_{3}$N &  3.7  &  0.50 & $-0.41$ ~h &  1.0  ~m \\
HC$_{3}$N &  2.5  &  1.3  & $-0.72$ ~m &  2.4  ~h \\
          &       &       &            &          \\
SiO       &  1.5  &  0.34 & $-0.61$ ~l &  0.49 ~m \\
SiO       &  4.2  &  0.20 & $-0.30$ ~m &  0.52 ~m \\
SiO       &  8.0  &  0.17 & $-0.00$ ~h &  0.51 ~m \\
SiO       &  4.5  &  0.50 & $-0.31$ ~m &  1.2  ~h \\
          &       &       &            &          \\
$^{13}$CS &  0.50 &  0.30 & $-0.66$ ~l &  0.17 ~m \\
$^{13}$CS &  1.5  &  0.15 & $-0.36$ ~m &  0.17 ~m \\
$^{13}$CS &  3.0  &  0.13 & $-0.06$ ~h &  0.17 ~m \\
$^{13}$CS &  1.5  &  0.37 & $-0.38$ ~m &  0.40 ~h \\
\hline
\end{tabular}
\end{center}
\end{table}

\section{Summary}
\label{sec:summ}

We have mapped a $2.5^{\circ} \times 0.5^{\circ}$ region of the Central
Molecular Zone (CMZ) of
the Galaxy using the Mopra radio telescope in 21 molecular and 3 hydrogen
lines emitting from 42--50\,GHz. The maps have spatial resolution 
$\sim 65$~arcsec
and spectral resolution 
$\rm \sim 1.8\,km\,s^{-1}$. 
For five species the
emission is particularly strong and widespread; CS, CH$_3$OH, HC$_3$N, SiO
and HNCO.  For the other molecules emission is generally confined to the
bright dust cores around Sgr A, Sgr B2 and G1.3.  These data add to, and
complement, previous studies we have undertaken in the 85--93 and
20--28\,GHz portions of the mm-spectrum. As also seen in these other bands
the molecular emission at 42--50\,GHz is both widespread, but asymmetric,
about the Galactic Centre, and turbulent.

Principal component analysis (PCA) of the ten brightest emission lines
quantifies the similarities and differences, with the first component, 
representing an
averaged integrated map of all the molecular lines, contributing 68\% of the
variance shown in the data.  The second component contributes 12\% to the
variance and highlights some differences in the dust cores, attributable to
the brightest lines being optically thick in them.  The third component
contributes 7\% to the variance and highlights some chemical differences,
with HOCO$^+$, HNCO and OCS being particularly prominent in G1.6 and Sgr~B2.

Line fluxes and luminosities are presented for the brightest dust cores, as
well as for the entire CMZ, to aid in comparison with observations of nuclei
of external galaxies where such components would be unresolved.  

Combining the data set with the higher-excitation lines we have mapped in
the 3\,mm band for five of the species (HNCO, HOCO$^+$, HC$_3$N, SiO and
$^{13}CS$) which indicates variations in the line ratios across the CMZ, with larger 
3-mm to 7-mm line ratios observed outside the $\rm 10^7\,M_{\odot}$ rotating dust ring, in particular for the molecular cloud complex G1.3.  This implies cooler temperatures for the molecular gas here than in the dust ring.

All these species are sub-thermally excited across the
CMZ, with excitation temperatures considerably less than the kinetic
temperature.  Modelling of the emission using RADEX suggests gas densities
for these molecules are, in general, a few $\rm \times 10^4 \, cm^{-3}$, a
factor of a few higher than corresponding densities derived from CO line
measurements.

The full data set, comprising the data cubes for the 21 molecular plus 3
hydrogen lines, is publicly available through
the Australia Telescope National Facility archives (http://atoa.atnf.csiro.au/CMZ).  A
final data set in this series, consisting of data on the isotopologues of
the CO J=1--0 lines over a $5^{\circ} \times 1.5^{\circ}$ of the CMZ, is
currently being prepared for publication.

\section*{Acknowledgments}
The Mopra radio telescope is part of the Australia Telescope National Facility 
which is funded by the Commonwealth of Australia for operation as a National 
Facility managed by CSIRO. 
The University of New South Wales Digital Filter Bank used for the observations 
with the Mopra Telescope was provided with support from the Australian Research 
Council (ARC).
We also acknowledge ARC support through Discovery Project
DP0879202.



\bsp

\label{lastpage}

\end{document}